\documentclass[12pt]{article}
\RequirePackage{amsthm,amsmath,amsfonts,amssymb}
\RequirePackage[authoryear]{natbib}
\RequirePackage[colorlinks,citecolor=blue,urlcolor=blue]{hyperref}
\RequirePackage{graphicx}
\usepackage{comment} 

\DeclareMathOperator*{\argmin}{arg\,min}

\title{Synthetic Control Misconceptions: Recommendations for Practice}
\date{\today}

\author{\thanks{New York University, Cash Transfer Lab}}

\author{Robert Pickett \thanks{New York University, Cash Transfer lab} \and Jennifer Hill \thanks{New York University, Department of Applied Statistics, Social Science, and Humanities} \and Sarah Cowan \thanks{New York University, Department of Sociology and Cash Transfer Lab}}
\date{\today}

\begin{document}

\maketitle


\begin{abstract}
To estimate the causal effect of an intervention, researchers need to 
identify a control group that represents 
what might have happened to the treatment group in the absence of that intervention. 
This is challenging without a randomized experiment and further complicated when few units (possibly only one) are treated. Nevertheless,
when data are available on units over time, 
synthetic control (SC) methods provide an opportunity to construct a valid comparison by differentially weighting control units that did not receive the treatment so that their resulting pre-treatment trajectory is similar to that of the treated unit. The hope is that this weighted ``pseudo-counterfactual" can serve as a valid counterfactual in the post-treatment time period. Since its origin twenty years ago, SC has been used over 5,000 times in the literature (Web of Science, December 2025), leading to a proliferation of descriptions of the method and guidance on proper usage that is not always accurate and does not always align with what the original developers appear to have intended. As such,
a number of accepted pieces of wisdom have arisen: (1) SC is robust to various implementations; (2) covariates are unnecessary, and (3) pre-treatment prediction error should guide model selection. We describe each in detail and conduct simulations that suggest, both for standard and alternative implementations of SC, that these purported truths are not supported by empirical evidence and thus actually represent \emph{misconceptions} about best practice. Instead of relying on these misconceptions, we offer practical advice for more cautious implementation and interpretation of results.
\end{abstract}



\section{Introduction}

In the twenty plus years since Abadie and Gardeazabal introduced synthetic control (SC) methods, researchers have used the approach thousands of times. In 2021, SC was cited by Guido Imbens as one of the most exciting prospects in econometrics in his Nobel Prize lecture \citep{guido_prize_2021}. The rapid rise in popularity of SC methods have inspired a proliferation of techniques and implementations that have outpaced the literature evaluating these approaches in applied settings. The primary support for these methods rests on proofs that posit particular data-generating processes, with surprisingly few evaluations of how these methods perform with finite pre-treatment observations and an unknown underlying data generating process.\footnote{For notable exceptions, see \citet{arkhangelsky_synthetic_2021, ferman_properties_2021, ferman_synthetic_2021, ferman_cherry_2020}.} 

Briefly, the goal of SC is to find a ``match" for a time series of a particular outcome variable prior to an intervention in a given treated unit. This `matched' pre-treatment trend is then extrapolated forward in time to the post-treatment period, which stands in for the hypothetical time series in the absence of treatment (the counterfactual). Because it is virtually impossible to find a single control unit that is a good match for the treated unit, we instead generate a `synthetic' match by taking a weighted average of untreated units on the basis of pre-treatment data. The causal effect estimate is then constructed as the difference between the outcomes for treated unit and the synthetic control unit. In practice, researchers tend to estimate causal effects that are the average of several time-specific treatment effects. 

Though the logic of SC is fairly straightforward, due to the generality of the theoretical guidance researchers face a number of practical decisions that may not have clearly defined best practices. This has caused some researchers to arrive at a few misconceptions about best practice in applied settings. 

\begin{itemize}
\item Misconception 1: SC is invariant to a number of implementation choices. This includes the choice of algorithm to generate SC weights and the choice of reference category for categorical or compositional covariates.
\item Misconception 2: Including covariates is unnecessary, especially if a close pre-treatment match for the outcome time-series can be found without them.\footnote{There are situations where including covariates is untenable, for instance when all pre-treatment outcome observations are included in the model \citep{kaul_synthetic_2015}. Here we consider the standard implementation of SC which summarizes the pre-treatment outcome series prior to estimation. Following \citet{ferman_cherry_2020}, researchers may wish to consider these results against a specification with all pre-treatment outcome observations and no covariates as a robustness check.}
\item Misconception 3: The closeness of the pre-treatment match can be used to adjudicate between different SC methods as well as different implementation choices within a given SC method.
\end{itemize}

In this paper, we conduct a simulation exercise to put these misconceptions to an empirical test. This paper proceeds as follows: we first outline an empirical case that motivates our exploration. Then, after a brief exposition of the inner workings of SC methods, we review these three misconceptions of SC implementation in more detail. We then present evidence from an empirically calibrated simulation study that challenges these misconceptions in practice. Finally, we conclude with recommendations for practice.

\subsection{Empirical Case: Alaska and the Permanent Fund Dividend}

To motivate our exploration, we draw on the empirical case of Alaska's Permanent Fund Dividend (PFD). In 1976, Alaska began investing a portion of its mineral revenues into a diversified fund in order to ensure long-term financial stability for the state. Since 1982 Alaska has paid out a portion of the fund's earned interest to eligible Alaskan residents as a dividend, with payments ranging from about \$1040 in 1984 to about \$3,800 in 2022 (in 2026 inflation adjusted dollars).\footnote{Payments in several other recent years, for example 2020 and 2025, were also on the order of \$1000.} This payment provides a per-family influx of cash roughly comparable to or exceeding payments from major social support programs including the earned income tax credit (EITC) and in kind payments from the supplemental nutrition assistance program (SNAP) \citep{cowan_examining_2022}. A number of papers have attempted to identify the causal effects of these annual cash transfers by comparing Alaska before and after 1982 on a number of outcomes. In particular, researchers have used SC methods to explore the PFD's effect on crime \citep{dorsett_bayesian_2021} and labor market participation \citep{jones_labor_2018}. To ensure our simulations are in line with practical applications, we use empirical data from Alaska to calibrate our simulation exercise (see Section \ref{simulation} for details).\footnote{We focus on this Alaska case study because it has been used in previous SC work and gives us a chance to ground our simulations in existing data. We are not, however, claiming that the issues raised in this paper are the only ones that might be raised with regard to the Alaska PFD. For instance, the creation of the fund is one of numerous major policy changes in Alaska in the wake of the oil boom, including major revisions to the tax code in 1975 and a complete repeal of the state income tax in 1980. This raises additional concerns about the volatility of the pre-treatment outcome series and anticipation effects due to multiple related policies being enacted over a short period of time just before the PFD was rolled out. Those issues have been discussed more thoroughly in the SC literature \citep{abadie_using_2021}. The purpose of this paper is to focus on the more insidious problems common across a broad range of potential SC applications.}

\section{Synth and its Implementation}

There are a number of statistical packages that estimate treatment effects using synthetic control methods, each with slight variations. Given the overwhelming popularity of the original work by Alberto Abadie and coauthors\footnote{The first reference to this approach as well as the Synth packages developed in Stata and R to implement it \citep{abadie_economic_2003} has been cited more than 7000 times and the seminal 2010 paper \citep{abadie_synthetic_2010} has been cited more than 8000 times. The Synth software has been downloaded more than one million times in Stata and over 180,000 times in R.} \citep{abadie_economic_2003, abadie_synthetic_2010} we use those packages as our starting point for describing the method. 

Assume we can write a factor model for an outcome $Y_{i,t}$ for unit $i$ in year $t$ as
\begin{equation*}
Y_{i,t} = \delta_t + \boldsymbol{\theta_t Z_i} + \boldsymbol{\lambda_t \mu_i} + \alpha_{i,t}D_{i,t} + \epsilon_{i,t} ,
\end{equation*}
where $\delta_t$ is an unobserved factor shared across units (year fixed effect), $\boldsymbol{Z}_i$ is a vector of $r$ observed covariates (observed once per unit pre-treatment), $\boldsymbol{\theta}_t$ is an unknown vector of $r$ time trends shared across units,  $\boldsymbol{\mu}_i$ is a vector of $f$ unobserved covariates (observed once per unit), and $\boldsymbol{\lambda}_t$ is an unknown vector of $f$ time trends shared across units. We will sort the units so the treated unit is the first ($i = 1$), followed by all control units ($i \in \{2, 3,  \dots n$\}). $D_{i,t}$ is a treatment indicator with values of 1 when $i = 1$ (the treated unit) and $t > T_0$ where $T_0$ is the last pre-treatment period (with $1 \leq T_0 < T)$, making $\alpha_{i,t}$ is the causal effect (defined below). Finally, $\epsilon_{i,t} $ is a vector of unobserved transitory shocks with mean zero. We will use $Y_{1,t}(1)$ and $Y_{1,t}(0)$ to denote the potential outcome within the treated state for the treated and control conditions, respectively, at time $t$, with $\alpha_{i,t} = Y_{1,t}(1) - Y_{1,t}(0)$. This set-up assumes that there are no treatment effects before the treatment starts, that the treated unit is continually treated after the treatment starts, and that the treatment has no effect on control units. 

The goal of synthetic control is to construct a synthetic counterfactual unit out of a weighted combination of control units, restricting these control unit weights to be convex (they are positive and sum to one) to avoid extrapolation. Specifically, if we find a vector of non-negative weights, $w_j^*$, for each control unit $j$, subject to $\Sigma w_j^* = 1$ such that 
\begin{equation*}
\sum_{j = 2}^{J+1}w_j^*Y_{j,t} \approx Y_{1,t}
\end{equation*}
for all $t \leq T_0$, and 
\begin{equation*}
\sum_{j = 2}^{J+1}w_j^*\boldsymbol{Z}_{j} \approx \boldsymbol{Z}_{1}
\end{equation*}
then $Y_{1,t}(0) - \sum_{j = 2}^{J+1}w_j^*Y_{j,t}$ will be bounded, small, and asymptotically approach zero as the number of pre-intervention periods gets large relative to the scale of the transitory shocks, $\epsilon_{i,t}$ \citep{botosaru_role_2019}. 

If these conditions hold, we can estimate the causal effect as, 

\begin{equation*}
\hat{\alpha}_{i,t} = Y_{1,t} - \sum_{j = 2}^{J+1}w_j^*Y_{j,t}
\end{equation*}

for $t \in \{T_0 +1, \ldots, T\}$.\footnote{\citet{abadie_synthetic_2010} also prove that the estimator works for an autoregressive model with time-varying covariates and coefficients,
\begin{equation*}
    \begin{alignedat}{2}
    Y_{i,t}(0) &= Y_{i,t} - \alpha_{i,t}D_{i,t}\\
    Y_{i,t+1}(0) &= \lambda_t Y_{i,t}(0) + \boldsymbol{\beta_{t+1}Z_{i,t+1}} + \mu_{i,t+1} \\
    \boldsymbol{Z_{i,t+1}} &= \gamma_tY_{i,t}(0) + \boldsymbol{\Pi_tZ_{i,t} + v_{i,t+1}}.
    \end{alignedat}
\end{equation*}
Both specifications assume shared trends in covariates $\boldsymbol{Z}$. We focus for now on the factor model for now as it more closely resembles their implementation but the autoregressive role will play a role in our simulation.} \citet{abadie_synthetic_2010} suggest that these weighting conditions may not hold exactly in practice, and it is up to researchers to evaluate ``if the characteristics of the treated unit are sufficiently matched by the synthetic control," without specific guidelines for how to make that determination. We will refer to the amount of pre-treatment mismatch between the synthetic control and treated units as the amount of `outcome imbalance,' and we will measure this imbalance with the pre-treatment root mean squared prediction error (RMSPE)\footnote{Other measures of imbalance are possible, and indeed different SC packages define different balance metrics. We use RMSPE here because it is common in the literature and in implementations of SC methods.} given by
\begin{equation*}
\sqrt{\dfrac{1}{T_0} \sum_{t=1}^{T_0} (Y_{1,t} - \sum_{j = 2}^{J+1}w_j^*Y_{j,t})^2}
\end{equation*}

In order to find the SC weights, we let $\bar{Y}_j^{\boldsymbol{K}_M}$ represent M linear combinations of pre-treatment outcomes $Y_{i,t}$ (usually just the state-specific mean prior to treatment, $\frac{1}{T_0}\sum_{t=1}^{T_0}Y_{i,t}$), and then let $\boldsymbol{X_1} = (\boldsymbol{Z_1}, \bar{Y}_1^{\boldsymbol{K}_M})$ for the treated unit, and $\boldsymbol{X_0}$ be the same for each of the $J$ control units, i.e., $\boldsymbol{X_0} = (\boldsymbol{Z_{i\neq1}}, \bar{Y}_{i\neq1}^{\boldsymbol{K}_M})$. We then want to choose weights $\boldsymbol{W}$ to minimize some distance between $\boldsymbol{X_1}$ and $\boldsymbol{X_0W}$. Because we cannot guarantee a match across all covariates, we want to use a vector of variable importance weights $\boldsymbol{V}$ to prioritize our matching. We can then define a variable loss function that reflects these priorities:
\begin{equation*}
\lVert \boldsymbol{X}_1 - \boldsymbol{X}_0\boldsymbol{W} \rVert_v = \sqrt{(\boldsymbol{X}_1 - \boldsymbol{X}_0\boldsymbol{W})'\boldsymbol{V}(\boldsymbol{X}_1 - \boldsymbol{X}_0\boldsymbol{W})}
\end{equation*}

To find weights $\boldsymbol{W^*}$ and $\boldsymbol{V^*}$ we can use a two-step optimization process.\footnote{There is some controversy here. Some scholars have suggested that this step should be approached explicitly as a bilevel optimization problem since the standard approach rarely finds optimal solutions (these optimal solutions are often corner cases where all weight is assigned to one predictor) \citep{malo_computing_2020}. Others have suggested disregarding $\boldsymbol{V}$ entirely, simply setting each value of the variable importance vector to 1 \citep{ben-michael_augmented_2021}.} The outer level will be a non-linear optimization that finds 

\begin{equation*}
\boldsymbol{V^*} = \argmin_v\frac{1}{T_0}\sum_{t=1}^{T_0}(Y_{1,t} - \sum_{j = 2}^{J+1}\boldsymbol{W^*(V)}Y_{j,t})^2.
\end{equation*}
The inner level of the optimization will be a quadratic optimizer that finds

\begin{equation*}
\boldsymbol{W^*(V)} = \argmin_w \lVert\boldsymbol{X}_1 - \boldsymbol{X}_0\boldsymbol{W}\rVert_v.
\end{equation*}

The nested optimizer is initialized with empirically derived $\boldsymbol{V}$ weights, which we will refer to as `regression weights.' This initial $\boldsymbol{V}$ vector is set by taking standardized summed squared regression coefficients for all covariates in $\boldsymbol{X}$ predicting pre-treatment outcomes in all years.\footnote{If the regression is inestimable, the nested optimizer is instead initialized with uniform $\boldsymbol{V}$ weights.} Specifically, let $\boldsymbol{X}$ be $\boldsymbol{X_1}$ appended to $\boldsymbol{X_0}$, $(\boldsymbol{X_0},\boldsymbol{X_1})$, and let $\boldsymbol{X_*}$ be $\boldsymbol{X}$ where all columns have been divided by their standard deviations, plus an intercept column. Then, $\boldsymbol{\beta}_{k,t} = \boldsymbol{(X_*'X_*)^{-1}(X_*'Y_{j,t})}$ for all covariates $k$ in $\boldsymbol{X}$ and all $t \in \{1, \ldots, T_0\}$. Finally, let the initial value for $\boldsymbol{V}$ be $\frac{\sum_t \boldsymbol{\beta}_{k,t}^2}{\sum_k\sum_t\boldsymbol{\beta_{k,t}^2}}$ (see also \citet{bohn_did_2014, kaul_synthetic_2015}).

While this empirically initialized nested optimization process is the default implementation used in R, it is not the default for Stata.\footnote{You can implement this two-step optimization in Stata by using the `nested' argument.} In Stata, the $\boldsymbol{V}$ matrix is simply set to the empirical estimate $\frac{\sum_t \boldsymbol{\beta}_{k,t}^2}{\sum_k\sum_t\boldsymbol{\beta_{k,t}^2}}$ and a single non-linear optimizer is used to find $\boldsymbol{W^*}$ conditional on the regression variable weights $\boldsymbol{V}$.

\section{Misconceptions of Synthetic Control}

Having reviewed the general set-up for SC methods, we now turn to a discussion of the three misconceptions we see guiding implementation of SC methods in practice.

\subsection{Misconception 1: Synthetic Control is Robust to Implementation Choices}
Synthetic control is now a common tool among applied researchers, who face many implementation decisions with unclear applied guidance. In this section, we discuss two such choices: which optimization process to use for variable weights, and which reference category to omit when including categorical or compositional covariates (or indeed whether to omit a reference category at all).

\subsubsection{Misconception 1a: Nested vs Regression Weights are Interchangeable}

\citet[pp. 496]{abadie_synthetic_2010} note that the proofs of Synthetic Control's asymptotic performance are valid for any choice of the variable weights vector $\boldsymbol{V}$. This can lead applied researchers to incorrectly assume that Synth is relatively invariable to choices for specific $\boldsymbol{V}$ vectors - thus treating the nested optimizer, `regression weights,' and a uniform $\boldsymbol{V}$ matrix as somewhat exchangeable. Differences in how SC is implemented across the two most popular statistical packages (`Synth' in R and Stata)  - without clear instructions for when to use one implementation or package over another - makes matters worse. R defaults to the nested optimizer and can only implement `regression weights' by manually inputting them as user-defined custom weights. Stata defaults to `regression weights' but can use nested weights by specifying the `nested' option.\footnote{The Stata documentation (\url{https://web.stanford.edu/~jhain/fqa.htm}) provides little guidance as to when you should use `nested' vs `regression' weights, simply suggesting `nested' weights as a strategy to improve pre-treatment fit.}

In their original paper, however, \citet[pp. 496]{abadie_synthetic_2010} also point out that the choice of $\boldsymbol{V}$ weights influences the mean squared error of the estimator, though we have not seen an empirical assessment of how substantial this influence might be. Thus, for our first test, we explore the degree to which SC's performance depends on choices of $\boldsymbol{V}$ in order to evaluate whether or not such decisions are reasonably inconsequential.

\subsubsection{Misconception 1b: Reference Category Choice is Inconsequential}

Next, we turn to the assumption regarding the choice of which reference category to exclude when including compositional or categorical covariates in SC. Categorical variables (e.g., race, gender, income bracket, marital status, etc.) often serve as important predictors of the outcome. For an example, \citet{jones_labor_2018} when estimating the effect of the Alaska Permanent Dividend on labor market participation, include three compositional variables: the percentage in age groups (with four categories), education groups (with three categories), and industry groups (with five categories).

When categorical or compositional variables are included in a linear regression, researchers typically omit one category as a reference category. No matter what category is omitted, the predictions from such a model would not be affected and the interpretations of all between group comparisons would be recoverable.

In contrast, the results of a standard synthetic control analysis \textit{are} sensitive to which reference category a researcher chooses. While a regression will encode the same information regardless of the reference category omitted, the sums of squared coefficient magnitudes, and thus regression variable weights, are not identical across these settings. For a specific example, consider one outcome variable $Y$, and mutually exclusive and completely exhaustive dichotomous regressors $X_1$, $X_2$, and $X_3$ such that $X_1 + X_2 + X_3 = 1$. If $E[Y|(X_1=1)] = 10$; $E[Y|(X_2 = 1)] = 1$; $E[Y|(X_3 = 1)] = -5$ then we can write equivalent regressions:

\begin{center}
$Y = -5 + 15*X_1 + 6*X_2 + \epsilon$

$Y = 1 + 9*X_1 - 6*X_3 + \epsilon$

$Y = 10 - 9*X_2 - 15*X_3 + \epsilon$

\end{center}

Though these regressions provide equivalent conditional expectations of $Y$, the sum of squared coefficients clearly varies: 261 in the first, 117 in the second, and 306 in the third. We could consider a similar example that includes another predictor $\alpha$ but keeps the coefficients for $X$'s the same: 

\begin{center}
$Y = -5 + 10*\alpha + 15*X_1 + 6*X_2 + \epsilon$

$Y = 1 + 10*\alpha + 9*X_1 - 6*X_3 + \epsilon$

$Y = 10 + 10*\alpha - 9*X_2 - 15*X_3 + \epsilon$

\end{center}

\begin{figure}[h]
    \includegraphics{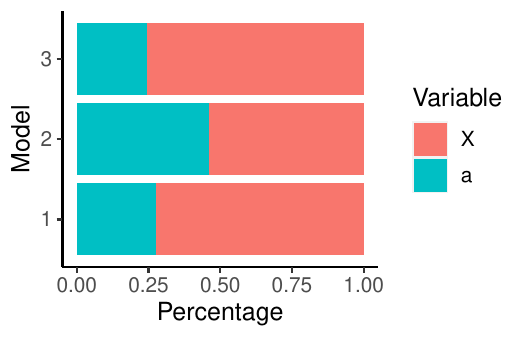}
    \caption{Contributions to $\boldsymbol{V}$}
\label{fig:ref_reg_weights}
\end{figure}

Though these regressions provide the same information (i.e., recover the same contrasts and produce the same predicted values), the choice of omitted category can change the implied importance of $\alpha$ relative to $X$ in the most popular version of the synthetic controls implementation. In the first equation, $\alpha$ makes up $\frac{10^2}{10^2 + 6^2 + 15^2} \approx$ 28\% of the total variable weight, in the second it makes up $\frac{10^2}{10^2 + 9^2 + 6^2} \approx$ 46\% of the weight, and in the third it makes up $\frac{10^2}{10^2 + 15^2 + 9^2} \approx$ 25\% of the weight (see Figure \ref{fig:ref_reg_weights}). In general, if we choose to omit a category where the conditional expectation of $Y$ for that category is farther away from the mean of conditional expectations of $Y$ for each possible omitted category we will increase the relative contribution of the categorical variable $X$ to the variable weights $\boldsymbol{V}$.

Imperfect matching between treated and synthetic units produces a second layer of uncertainty that is independent of the $\boldsymbol{V}$ matrix. The standard logic is that if you match on all but one category of a compositional or categorical variable, the linear dependence among categories will ensure that you also match on the category left out. Unfortunately, the curse of dimensionality makes finding a perfectly matching convex combination of control units impossible in many applied settings. 

In practice, many imperfect solutions are possible, and the specific solution will depend on which categories are included in the analysis. Thus, selecting different reference categories will produce slightly different synthetic control weights that in turn yield slightly different matches between the treated and synthetic control unit. These different matches produce different causal estimates.

\subsection{Misconception 2: Covariates are not necessary when using SC}

In their proof for when the synthetic control approach can be unbiased, \citet{abadie_synthetic_2010} show that if the SC assumptions hold the bias introduced by omitting covariates goes to zero as the number of pre-treatment time periods grows large relative to transient shocks in the outcome (see also \citet{botosaru_role_2019, ferman_synthetic_2021, kaul_synthetic_2015}). The intuition behind this result is that because the outcome series is of primary interest, covariates only matter to the extent that they influence the outcome time series, and if the Synthetic Control method can perfectly reproduce the untreated outcome time series, additionally matching on covariates is superfluous. In other words, if you can closely align the treated and synthetic outcome trajectories for a long enough time before treatment --- and the data generating process follows either a factor model or a specific autoregressive model \citep{abadie_synthetic_2010} --- you must also have aligned on the factors that are relevant for producing that outcome. This has led to some to de-emphasize the importance of covariates in SC analyses \citep{botosaru_role_2019, gilchrist_2023}. Others have recommended excluding covariates in some circumstances for practical reasons, e.g., when opting to balance on all pre-treatment outcomes, which may also reduce variation in results across plausible implementations \citep{ferman_cherry_2020}.

We offer two words of caution for those considering omitting covariates. First, covariates may have substantial effects on the performance of the estimator, especially when the pre-treatment time series is short \citep{kaul_synthetic_2015}. Second, the pre-treatment fit diagnostic may be an unreliable measure of how closely the synthetic control resembles the counterfactual when the pre-treatment time series is short \citep{abadie_using_2021}. Consider an extreme example where we have observations for a single time point prior to treatment and the treated unit's data absent treatment can be written as:
\begin{equation*}
Y_{1,t}(0) = 1 + \epsilon_{1,t}; \epsilon_{1,t} \sim N(0, 1)
\end{equation*}

and data for all control units can be written as:

\begin{equation*}
Y_{j,t} = 0 + \epsilon_{j,t}; \epsilon_{j,t} \sim N(0, 1)
\end{equation*}

In this scenario, if we have a sufficient number of control units it is quite likely that we can find a perfect match for the treated unit's single pre-treatment observation. That said, we've constructed this perfect match by selecting control units with randomly large values of $\epsilon_{j,1}$ to match the structurally large values of $Y_{1,1}$. We've effectively overfit our model to transient shocks rather than structural variation in the outcome series. When we extrapolate the post-treatment synthetic control time series, the expected value of $\sum_{j = 2}^{J+1}w_j^*Y_{j,t}$ will return to zero, while the expected value of the true counterfactual $Y_{1,t}(0)$ will remain 1. Thus, whatever causal estimate we generate will be upwardly biased by 1 unit in expectation. In sum, we can have an exact empirical match while having a remarkably poor match in expectation. When there are few pre-treatment periods, including relevant covariates may reduce overfitting by providing additional information. We explore whether the logic of this toy example holds in general practice in our simulation results.

\subsection{Misconception 3: Lower Pre-Treatment Outcome Imbalance Suggests Lower Absolute Bias}

In their original papers describing the method \citet{abadie_comparative_2015, abadie_synthetic_2010} note that if pre-treatment outcome imbalance is poor, synthetic control methods are unlikely to produce unbiased estimates of the treatment effect. Over time, though, this general caution seems to have been interpreted as a recommendation to use pre-treatment outcome imbalance as a metric for model selection. For example, \citet{Panagiotoglou_2022} and \citet{oliphant_2022} restrict the years they use for pre-treatment opimization to improve their pre-treatment outcome imbalance. Overemphasizing the goodness of fit metric in this way is likely to lead to overconfidence in the model results and potential overfitting to idiosyncratic changes in the outcome variable over short durations. An alternative interpretation of this strategy of omitting pre-treatment data to improve goodness of fit is to view the exercise as a failed pre-treatment placebo test, as it suggests that the synthetic control model is not providing reliable pre-treatment predictions across all available data. 

Others may place too much emphasis on pre-treatment outcome imbalance in subtler ways. For example,  \citet{gilchrist_2023} and \citet{donohue_2019} tie the plausibility of synthetic control estimates to low pre-treatment imbalance. Although results from synthetic control models that produce large deviations from the pre-treatment time series should be viewed skeptically, closely reproducing a short pre-treatment time series is not sufficient for trustworthy results. Indeed, over short durations these models can perfectly fit idiosyncratic noise, producing inaccurate estimates of the future time series. \citet{zimmerman_californias_2021} and \citet{townsend_use_2022} go further, relying on pre-treatment mean squared prediction error to determine whether or not they include covariates in their analyses. Balancing on covariates provides the opportunity for additional verification in the case that models fit to short pre-treatment outcomes are in fact providing reasonable estimates of the structural variation for the treated unit. Simply discarding them when the covariates suggest worse fits ignores this potential warning. 

Others rely on pre-treatment balance to guide model fitting decisions. \citet{opatrny_impact_2021} determines which control units to include by examining which set produces the lowest pre-treatment RMSPE, and \citet{islam_local_2019} and \citet{propheter_effect_2020} rely on pre-treatment RMSPE for variable selection --- including the set that produces the lowest pre-treatment prediction error. Ideally, these decisions would instead be made according to theories about the pre-treatment data generating process, or by following a set procedure to minimize potential bias from specification searching \citep{ferman_cherry_2020}.

Relatedly, \citet{Parast2020} develop metrics for identifying when pre-treatment balance is sufficient for accepting the synthetic control estimate - but also note that minimizing imbalance may not always produce the most plausible estimate (e.g., using fewer pre-treatment periods). While the intuition behind using pre-treatment RMSPE in these ways is clear, we have not seen this relationship explicitly tested in the literature. In the results below, we explore whether pre-treatment outcome imbalance is predictive of post-treatment model performance, or whether researchers should only rely on pre-treatment outcome imbalance to identify when synthetic control methods are failing entirely.

\section{Simulation} \label{simulation}
To better understand the potential effects of these misconceptions we conducted a simulation analysis to compare the performance of a variety of approaches relative to a known truth. We ground our analysis in a facsimile of an observed empirical scenario to better understand how these misconceptions may play a role in practice. 

The empirical setting we use to calibrate our simulation focuses on the effect of the 1982 Alaskan Permanent Fund Dividend. In particular, we use data from the Current Population Survey (CPS) to estimate pre-treatment relationships between covariates and state-specific trends and base our simulated data on these estimates. We outline this process in the next sections and provide further details in Appendix \ref{appendix:sim}.

\subsection{Simulation Overview} \label{simulation.settings}
Our simulation is designed  to explore the tenability of the three  misconceptions described above. As such it is designed with three factors, each of which controls either aspects of the data generating process or implementation choices that could impact performance of the treatment effect estimators.

Since the original implementation of SC appears to rest heavily on the no-extrapolation assumption, our first simulation factor creates a range of scenarios that either fully support or violate that assumption. As such this factor creates variation in the researcher's ability to create a synthetic control that approximates the treated unit prior to treatment exposure. We achieve this by varying the empirical overlap between the treated and control states when generating the covariate distributions. 

The second simulation factor is implemented in two different models for the data generating processs (DGP) for the outcome. The first is a factor model that closely aligns to the proofs laid out by \citep{abadie_synthetic_2010}. The second is a linear model that resembles standard regression analyses, but for which there is no clear asymptotic proof as it relates to SC. 
The third simulation factor explores model performance across omitted reference categories for compositional or categorical variables. 

We evaluate the performance of four methods (each of which is implemented in several ways) across these three simulation factors. 
These four methods vary in the degree to which they allow for extrapolation from control units in order to fit the treated outcome series. In order from least extrapolation to most extrapolation, we test Synth, Augsynth, GSynth, and Bayesian Structural Time Series (BSTS). In addition, we test the performance of each of these methods with and without covariates.

\begin{figure}[h]
    \caption{Simulation Overview}
    \includegraphics[width=\textwidth,height=\textheight,keepaspectratio]{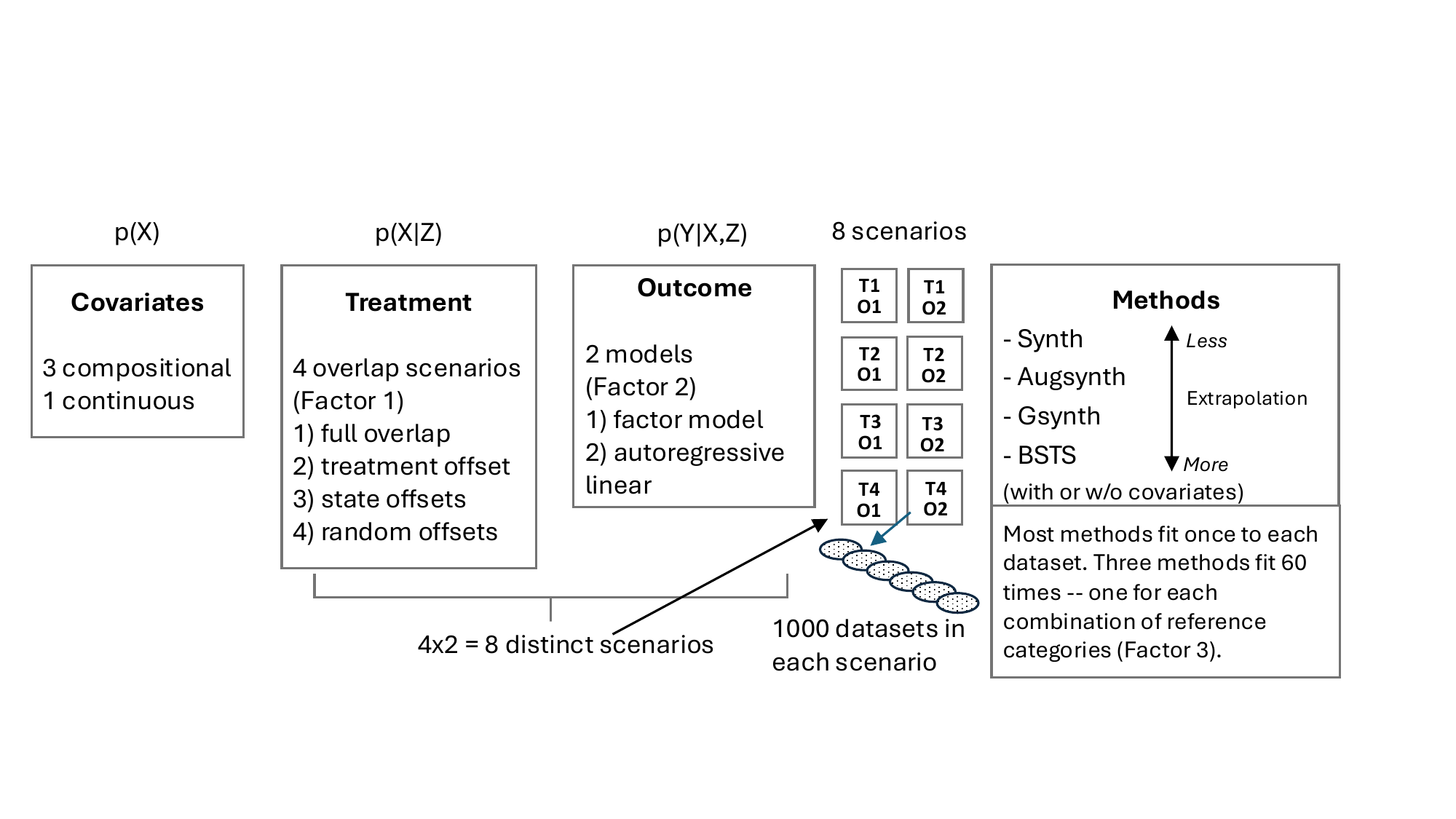}
\label{fig:SimConditions}
\end{figure}

An overview for this simulation setup can be found in Figure \ref{fig:SimConditions}. We describe each of these steps in the following sections and additional detail can be found in Appendix \ref{appendix:sim}. In Appendix \ref{appendix:pre} we additionally investigate model performance for our simulation settings across three different pre-treatment time series durations.

\subsection{Simulation Calibration}

We calibrate our simulations to ``real life" by fitting models to 10 years of data from the Current Population Survey Annual Social and Economic Supplement (CPS-ASEC). We then use estimates of those model parameters to simulate 100 years of new data. Because we are primarily interested in the Permanent Fund Dividend (PFD), we use the 10 years of data surrounding the implementation of the PFD, 1977--1986.\footnote{1977 was the first year Alaska was included in the CPS-ASEC.} Our outcome variable of interest is the proportion in the state who are employed part-time. The average proportion working part-time across all states in the CPS between 1977 and 1986 is 9.5\% with a standard deviation of 1.7\%. For Alaska specifically, the mean is 8.4\% with a standard deviation of 0.9\%. In our models we include covariates for the following: the racial composition of states (percent White, Black, and Other, which sum to 100\%); the educational composition of states (percent with less than high school education, high school, some college, and college or more); the composition of people working in five industry categories (the percent working in agriculture, forestry, fisheries, mining, construction, and manufacturing; the percent working in transportation, communications, other public utilities, wholesale trade, or retail trade; the percent working in finance, insurance, real estate, business and repair services, and personal services; the percent working in entertainment and recreation services, professional and related services, public administration, active duty military, and experienced unemployed not classified by industry; and the percent not currently in the labor force; and the average self-reported annual wage by state. The inclusion of these three compositional covariates generates 60 possible combinations of reference categories.\footnote{Three race categories by four education categories by five industry categories produces 60 possible reference category combinations.}

 Our observed time series only spans 10 years (5 years pre-treatment and 5 years post-treatment), but we need sufficient data to estimate synthetic control methods. Therefore, we "artificially extend" our simulated data. To do this, we pretend that our 10 years of data actually spanned 100 years (50 pre-treatment and 50 post-treatment).\footnote{While studies with one hundred years of data may not occur often, recall that synthetic controls can be used in a wide variety of circumstances with time trends. For instance, the data could be measured in minutes, hours, days, or weeks. In those circumstances, 100 pre-treatment "time periods" could easily occur.} Therefore we relabel observed year 1977 as our $0^{th}$ year in our simulation world (50 years pre-treatment) and 1986 as our $100^{th}$ year in our simulation world (50 years post-treatment). Each one year gap in the real world then corresponds to a 10 year gap in the simulated world. This means that our simulation model relies on \emph{interpolations} between observed years rather than \emph{extrapolations} beyond the range of our observed covariate data. This helps to ensure that the covariate data used reflect conditions that are likely to exist in the world.

\subsubsection{Simulating Covariates and Inducing Treatment Assignment}

To build the covariate portion of our data generating process while calibrating to the Alaska data, we start by considering a factorization of the joint likelihood of the covariates into the following distributions: the conditional distribution of industry given state and year; the conditional distribution of education given industry, state, and year; the conditional distribution of race given industry, education, state, and year; and the conditional distribution of wage given the rest. The first three conditional distributions are modeled as Dirichlet regressions and the last as a standard linear regression with normal errors. Sampling predicted values from these models and their associated uncertainty parameters will allow us to preserve observed relationships between covariates in our simulated data.

The strength of synthetic control methods lies in their ability to reconstruct synthetic counterfactual units that are plausibly similar to the treated unit prior to treatment. It is crucial, then, to evaluate how well these models can recover these plausible counterfactuals when treated units are very similar to donor control units, versus when treated units are less similar. Recall that SC doesn't require any one comparison unit to be very similar to the treated unit but rather that a weighted combination of the comparison units is similar to the treated unit. If the comparisons states are systematically different, then this weighted average will require extrapolation beyond the observed covariate space for the controls in order to create a pseudo-control that is sufficiently similar to the treated unit.

To evaluate the performance of these models under different conditions, we generate data from a range of scenarios that vary based on the extent to which our treated state (Alaska) is more or less similar in its covariate values to the 49 other potential donor control states. We operationalize these differences by fitting models with varying fixed effects structures to the observed data, which then inform the models from which we generate simulated data.

We construct four such scenarios:

\begin{itemize}
    \item[1.] In the \emph{full overlap} scenario we omit all state fixed effects from our models thus the structural variation in pre-treatment covariates for treated and control units is forced to be identical. 
    \item[2.] In the \emph{treatment offset} scenario the pre-treatment covariate models allow for a distinct intercept only for the treated unit (i.e. it alone has a state fixed effect). 
    \item[3.] In the \emph{state offset} scenario the pre-treatment covariate models include a state fixed effect for each state. This is the closest scenario to the empirically observed overlap.
    \item[4.] In the \emph{random offset} scenario the models include a state intercept randomly drawn from a shared distribution. By randomly assigning state intercepts we allow for idiosyncratic differences between states within any given simulation dataset, but set the expectation of those differences to be zero across the full set of simulations. 
\end{itemize}

To simulate covariate data, we first fit each model based on the empirical data and model specifications described above. We then use the parameter estimates as means when drawing coefficients for each regression model (drawn from a multivariate normal distribution) that simulates 100 years of data. We can then draw from a Dirichlet (or normal) distribution conditional on these parameters and the observed data. We provide further details in Appendix A.

\subsubsection{Simulating outcomes and treatment effects}
After generating simulated covariates, our next step is to simulate outcomes, conditional on the covariates. As our second simulation factor, we simulate outcomes in each of two different ways to reflect slightly different assumptions about the nature of the data generating process. One approach assumes an autoregressive linear model and the second assumes a factor model (consistent with the original SC assumptions). As with the covariate generation, we start by fitting the assumed model to Alaska.

To avoid incorporating any effects of the true PFD policy on the outcome, we only use CPS-ASEC (Alaska) data from years 1977 through 1981 to fit these outcome models. For the outcome models we do not incorporate any systematic variation across states. Thus, in our four covariate scenarios, the outcome model is the same for all states; therefore, the differences between states with regard to outcome variables will only occur in situations where their covariates have been modeled to be systematically different. In this sense, these simulations provide a generous testing ground for synthetic controls because if there is overlap with respect to covariates there should also be overlap with respect to outcomes.

For the autoregressive model, we regress the proportion working part-time on compositional industry, education, and race variables as well as our continuous average wage variable.\footnote{Though the proportion working part-time is constrained to be within the range [0,1], we fit linear regressions for simplicity.} For the factor model, we first calculate state-specific pre-treatment means for each of our covariates in the CPS data. We then regress the proportion working part time on year, these state-specific pre-treatment means, and the interaction between year and these pre-treatment means. As with the covariates, we generate simulated outcome variables by drawing coefficients from their estimated multivariate normal distribution and then drawing from a normal distribution conditional on those parameter values and the data. 

Finally, for each outcome simulation model, we add a constant treatment effect for simulation years $\geq 50$ in the state of Alaska equal to two times the observed standard deviation of the proportion working part-time in Alaska in the CPS-ASEC for years 1977 through 1986.\footnote{The constant treatment effect is $\approx 1.7\%$. Because Synthetic Control Methods are fit using exclusively pre-treatment data, the magnitude of the treatment effect has no effect on the $\boldsymbol{w^*}$ weights, and, in turn, the estimated counterfactual trajectory.} Our simulations will examine how accurately each method can replicate this ground truth.

In all, the two types of outcomes (factor two) generated for each of four covariate scenarios (factor one) produce eight distinct simulation scenarios. To make comparisons as clear as possible, covariate values are identical for each pair of outcomes within the four covariate scenarios. We conduct this simulation 1,000 times, producing 8 outcome (factor vs autoregressive) by covariate (``full overlap", ``treatment offset", ``state offset", and ``random offset") scenarios each with 1,000 simulated datasets. 

Within each of the 8 simulation scenarios that capture variation in simulation factors 1 and 2, we perform a test of the implications of the variation induced by factor 3, which focuses on reference category choice. In particular, for each of these datasets, we generated average causal estimates for all 60 possible reference category combinations for methods that vary in the ways that they handle such covariates (described in the next section). The resulting uncertainty across these 60 models then gives us an assessment of how sensitive these synthetic control causal estimates are to reference category choice. 

\subsection{Method Details}
Overall we tested twelve approaches. Nine of these implementations generate consistent estimates regardless of omitted reference categories;  three do not (described below) and thus were used in our examination of the implications of reference category choice (factor 3). We select this set of implementations because they provide a range of assumptions about the data generating process as well as a range of approaches to extrapolation. We briefly outline the four categories of methods included as well as the specific implementations used below.

\subsubsection{Standard Synthetic Control}
The Synth models used follow the details laid out in \citep{abadie_synthetic_2010} described in this paper's introduction. Covariates can be included into the analysis using a nested optimization procedure that balances treated and control units on covariates, weighted for how influential those covariates are in predicting the outcome. Both these variable importance weights and donor weights are constrained to be between 0 and 1, and sum to one, requiring that the synthetic control unit be on the convex hull of control units. In other words, this method can provide an interpolation of control units but cannot extrapolate beyond them.

We use four different implementations of Synth. The first two are important for the exploring the implications of reference category choice (factor 3) because the estimates are sensitive to this choice. These are (1) standard \texttt{Synth} with nested optimizer weights \footnote{The nested optimizer is initialized with regression weights by default and is sensitive to initial values. Therefore by default the reference category still influences the weights produced by the nested optimizer.} and (2) standard \texttt{Synth} with regression weights.

While exploring the implications of the other two simulation factors we include two additional method implementations that are invariant to reference category choice. The first is Synth with all categories included.\footnote{This must use the nested optimizer, and relies on a uniform initialization for $\boldsymbol{V}$.} The second is Synth with no covariates.

\subsubsection{Augmented Synthetic Control}
Augsynth \citep{ben-michael_augmented_2021} augments this standard Synth estimator by including a ridge regression predicting post-treatment outcomes among control units with pre-treatment outcomes and covariates as predictors. This regression component allows regularized extrapolation from the convex hull, when it is untenable to construct a plausible counterfactual through interpolation alone.\footnote{Augsynth also adjusts the SC procedure for determining synthetic control weights by incorporating a dispersion penalty in the estimation and setting variable importance weights, $\boldsymbol{V}$, to the identity matrix.} This augmentation allows for limited extrapolation from the convex hull of control units to improve pre-treatment fit. 

Residualized Augsynth modifies the Augsynth estimator by setting the ridge penalty parameter to zero for covariates, but keeping it greater than 0 for pre-treatment outcomes, creating augmented synthetic control weights that perfectly match on auxiliary covariates. Because these weights now perfectly match on covariates, it no longer matters which reference categories we omit --- as a perfect match for $k-1$ categories of a compositional variable implies a perfect match for the $k$\textsuperscript{th} category as well.

To explore variation due to reference category choice we use generic \texttt{Augsynth}. Augsynth uses a fixed, constant, $\boldsymbol{V}$ matrix, which means the uncertainty due to reference category choice can only arise from imperfect matching across covariates. We also include two methods that are invariant to reference category choice. The first is Augsynth with all categories included. The second is Augsynth with no covariates.

\subsubsection{Generalized Synthetic Control}
The generalized synthetic control (GSynth) approach \citep{xu_generalized_2017} recasts the modeling problem as a linear interactive fixed effects model. This model writes the outcome as a function of time-varying covariates, a time-varying treatment effect, and the interaction of state and year fixed effects. Because the regression coefficients are unconstrained, this approach does not restrict the amount of extrapolation from the convex hull of control units. The time-varying treatment effect is estimated by first estimating regression coefficients for the time-varying covariates, year fixed effects, and control state fixed effects on data from only the control data. The treatment state fixed effect is estimated by taking the mean of the pre-treatment outcome series for the treated state after subtracting the time fixed effects and covariate effects estimated from the control unit data. 

The treatment effect, then, is estimated by the difference in the post-treatment outcome series for the treated state and a counterfactual outcome series constructed by recombining the effects of time-varying covariates and the interaction of state and year fixed effects. Because this approach incorporates covariates by directly regressing them on the outcome, and as noted earlier, which category of a compositional or categorical variable is omitted as a reference category is inconsequential for predicted values, this generalized synthetic control method is insensitive to reference category choice. We implement two versions of GSynth, one that includes covariates and one that does not.

\subsubsection{Bayesian Structural Time Series}
Finally, Bayesian structural time series (BSTS) \citep{brodersen_inferring_2015}, reframes the problem as another type of regression model --- in this case a state-space model. This approach also allows for unlimited extrapolation from the convex hull, but presumes a different data generating process from GSynth. This state-space model is composed of any or all of 1) a local linear trend, 2) a seasonality component, and 3) the effects of contemporaneous covariates. This model sets the outcome as only the treated state's outcome series. The covariates used to predict this outcome series include the treated state's contemporaneous covariates, the outcome series for control states, and the contemporaneous covariates from control states. For Synthetic Control applications, the state-space model is constructed using only data prior to treatment. Post-treatment projections from this state-space model are then treated as counterfactual comparisons for the true post-treatment data for the treated state. 

This model has fifty pre-treatment observations for the treated state as an outcome, and up to 713 covariates in the regression (51 states times one outcome series plus one continuous covariate plus three race covariates plus four education covariates plus five industry covariates then subtracting one treated state outcome series). Such a regression requires heavy regularization in order to be estimable. In the case of BSTS, this is done through a spike-and-slab prior on all covariates that effectively ``turns off' most covariates in the model. Because this approach is so highly regularized, we included all categories for our compositional variables and let the model ``choose" which to keep as part of its larger regularized variable selection process. We include two versions of this model with respect to covariates. One includes all covariates\footnote{Of the four methods explored, Augmented Synthetic Control and Bayesian Structural Time Series incorporate regularization in variable selection, allowing those models to converge despite not omitting collinear reference categories. Synthetic Control with a nested optimizer (and uniform initialization) does not directly rely on regression and can similarly incorporate perfectly collinear predictors. Only Generalized Synthetic Control requires omitting reference categories for convergence, but these regression models produce output that are insensitive to reference category choice.} and one includes none.
    
\section{Results}
We organize our simulation results by the misconceptions described above. 

\subsection{Misconception 1: Synthetic Control is Invariant to Implementation}
\begin{figure}[h]
\centering
\includegraphics[width=\textwidth,height=\textheight,keepaspectratio]{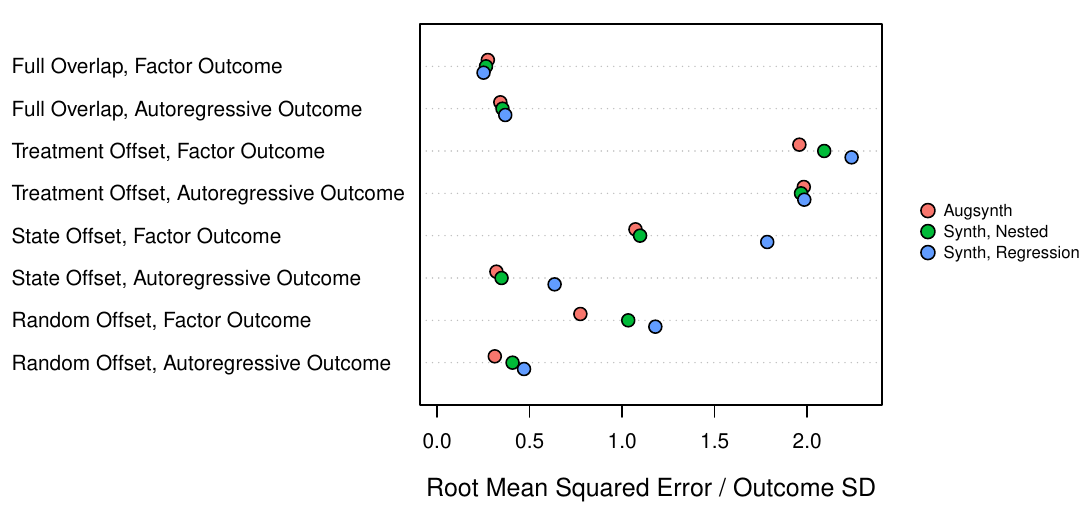}
\caption{\label{fig:fig1}Average Root Mean Squared Error of the estimated causal effect by Simulation Scenario, Method, and Outcome Data Generating Model. Vertical jitter added to help distinguish between points representing methods with similar performance.}
\end{figure}

We start by exploring the variation in performance of different synthetic control specifications at a high level, focusing first on the aggregate differences between Augsynth, Synth with a nested optimizer (R's default), and Synth with regression weights (Stata's default), the primary implementation choices for SC. Figure \ref{fig:fig1} displays the average root mean squared error of the estimated causal effect across all reference category combinations for all simulation scenarios for each of these methods.

This results suggests that there are several scenarios where there is no clear difference in average RMSE of the estimated causal effect across methods --- particularly both ``full overlap" scenarios, and the ``treatment offset" scenario with an autoregressive outcome model. However, in three scenarios (both ``random offset" scenarios and the ``treatment offset" scenario with a factor outcome model), there is a clear progression in performance, with Augsynth having the lowest average RMSE, followed by Synth with nested weights, and finally Synth with regression weights having the highest average RMSE. In the two ``state offset" scenarios, there is little difference in the average RMSE between Augsynth and Synth with nested weights, but Synth with regression weights has substantially higher average RMSE. In all, when it comes to average RMSE, there appears to be a clear ordering across all simulations --- Augsynth tends to perform the best and Synth with regression weights tends to perform the worst. The only scenarios where this ordering doesn't hold, differences in average RMSE across methods are negligible. \emph{We thus find that the choice of synth implementation can lead to substantial variation in the outcomes} - especially when there is not perfect overlap between treatment and control units.

\begin{figure}[h]
\centering
\includegraphics[width=\textwidth,height=\textheight,keepaspectratio]{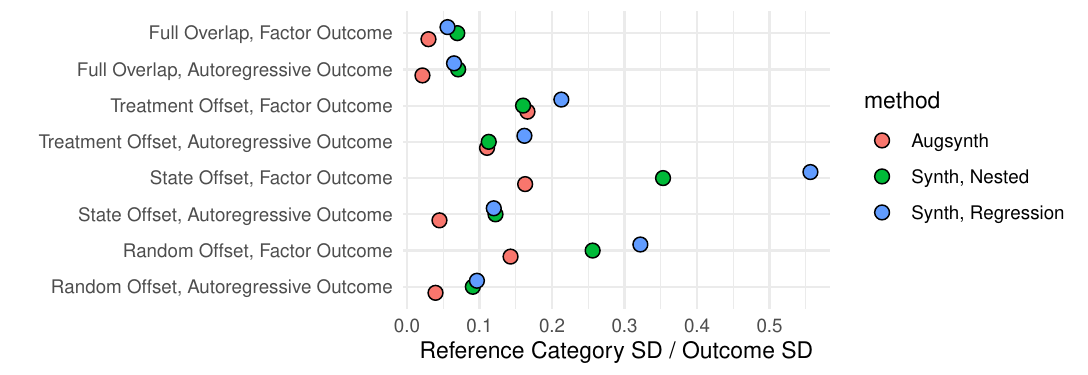}
\caption{\label{fig:fig2}Variation in Synthetic Control Estimates Due To Reference Category Choice by Method and Simulation Scenario. Vertical jitter added to help distinguish between points representing methods with similar performance.}
\end{figure}

Figure \ref{fig:fig2} explores whether reference category choice (factor 3) is truly arbitrary and thus has little to no effect on the resulting estimate. For each of the same set of methods used in the previous figure and for the eight simulation scenarios, this plot displays the standard deviation of the average treatment effect estimates across all sixty possible reference category combinations as a fraction of the overall standard deviation in the outcome variable.

\emph{The seemingly arbitrary choice of a reference category can markedly change the results.} The amount of variation in estimates across reference categories is larger for both of the Synth methods, ranging from just over 5\% of the outcome standard deviation in the ``full overlap scenarios" to 35\% and 55\% for nested and regression weights respectively in the ``state offset" scenario with a factor outcome. Contrary to the assumption that the choice of reference categories is inconsequential for the resulting ATT estimate, we find substantial variation in estimates across reference categories for all methods in all scenarios, with the most variation in Synth with regression weights estimates and the least variation in Augsynth estimates. While Augsynth is the least variable overall, in some scenarios (``state offset", ``random offset", and ``treatment offset" with factor outcome models) Augsynth can still have variation upwards of 15\% of the outcome standard deviation.

\subsection{Misconception 2: Covariates are not necessary}
\begin{figure}[h]
\centering
\includegraphics[width=\textwidth,height=\textheight,keepaspectratio]{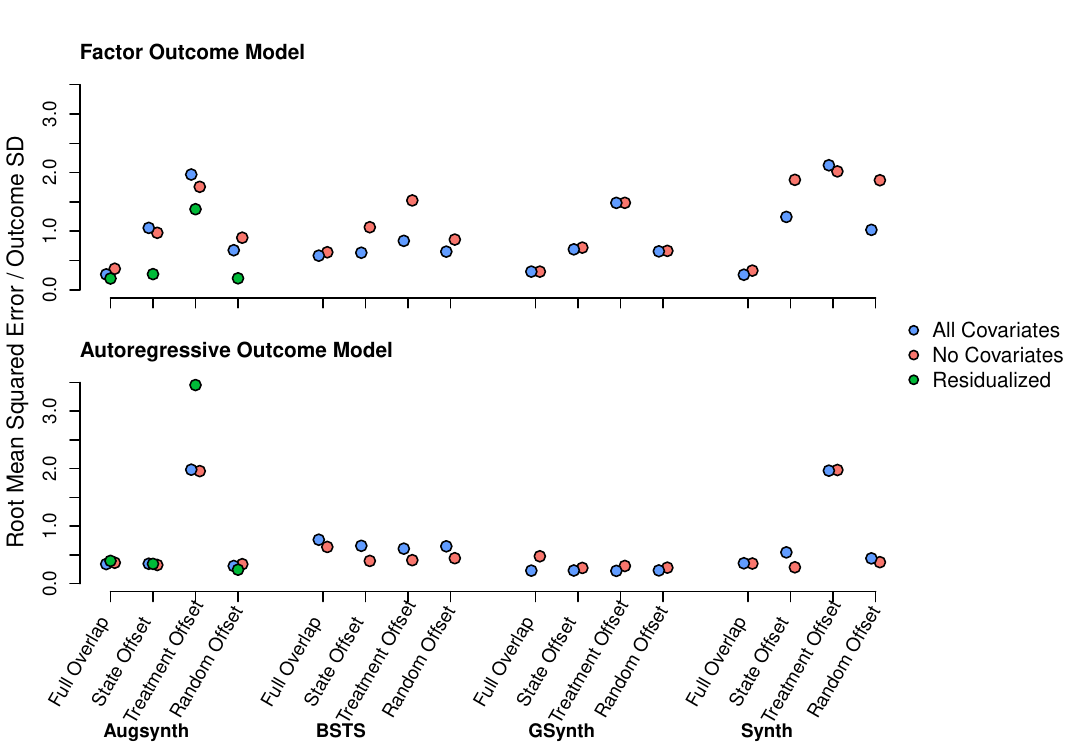}
\caption{\label{fig:fig3}Root mean squared error of the estimated causal effect using All or No Covariates by simulation scenario, method, and outcome data generating model. Horizontal jitter added to help distinguish between points representing methods with similar performance.}
\end{figure}

We next explore the belief that covariates are superfluous to synthetic control methods, and that they can be reasonably ignored with little consequence for the resulting estimates. Figure \ref{fig:fig3} shows the RMSE of the estimated causal effect for Augsynth, BSTS, GSynth, and Synth across simulations for each scenario as a function of how each methods handles covariates (e.g. including them or excluding them entirely). 

We include all covariates as an option to give each method the best chance to fit the data (i.e., limitations in model performance are not due to us arbitrarily excluding crucial information). Our estimation choices provide a sense of the full range of variation in implementation with respect to covariate inclusion, but we imagine that applied researchers will often find themselves somewhere in between these extremes.

Figure \ref{fig:fig3} displays results separately for implementations without covariates, with all covariates, and residualized for the Augsynth implementation. For BSTS and Synth, we compare implementations without covariates and with all covariates included, and for GSynth we compare implementations without covariates and with all covariates and arbitrary reference categories.\footnote{GSynth produces the same estimate regardless of what reference category is omitted. For the Synth implementation with all covariates included, we start with a uniform initialization of the V matrix and bypass the regression which would fail due to perfect collinearity among predictors.} Within the Augsynth methods, residualization tends to produce lower RMSE of the estimated causal effect, with the exception being the ``treatment offset" scenario with an autoregressive outcome model --- in which both the all covariates and no covariates implementations produce similar RMSE that is lower than the RMSE for residualization. For both Synth and BSTS, including all covariates tends to reduce the RMSE for the factor outcome model, but increase RMSE for the autoregressive outcome model (though the amount of increase for the autoregressive outcome model is less than the reduction for the factor outcome model). Within GSynth, implementations that include all covariates either have lower or the same RMSE. \emph{Overall, the relationship between the inclusion of covariates and root mean squared error of the estimated causal effect is ambiguous, but including covariates or residualizing tends to produce lower RMSE than excluding covariates entirely.}

To further disentangle the relationship between the inclusion of covariates and RMSE of the estimated causal effect, we include a commonly used measure of how well the synthetic state aligns with the treated state's pre-treatment outcome series. We'll refer to this measure as `Outcome Imbalance' and calculate it as the mean squared difference between the treated state's outcome trajectory and the synthetic state's outcome trajectory prior to treatment./footnote{This is often described as `Root Mean Squared Prediction Error' in the literature, but we refer to 'outcome imbalance' to better distinguish from the root mean squared error of the estimated causal effect.} To explore the hypothesis that the ambiguous relationship between RMSE and covariates might be due to differences in pre-treatment fit, we compare the average absolute bias of methods with and without covariates conditional on whether including covariates improved pre-treatment outcome imbalance.

\begin{figure}[h]
\centering
\includegraphics[width=\textwidth,height=\textheight,keepaspectratio]{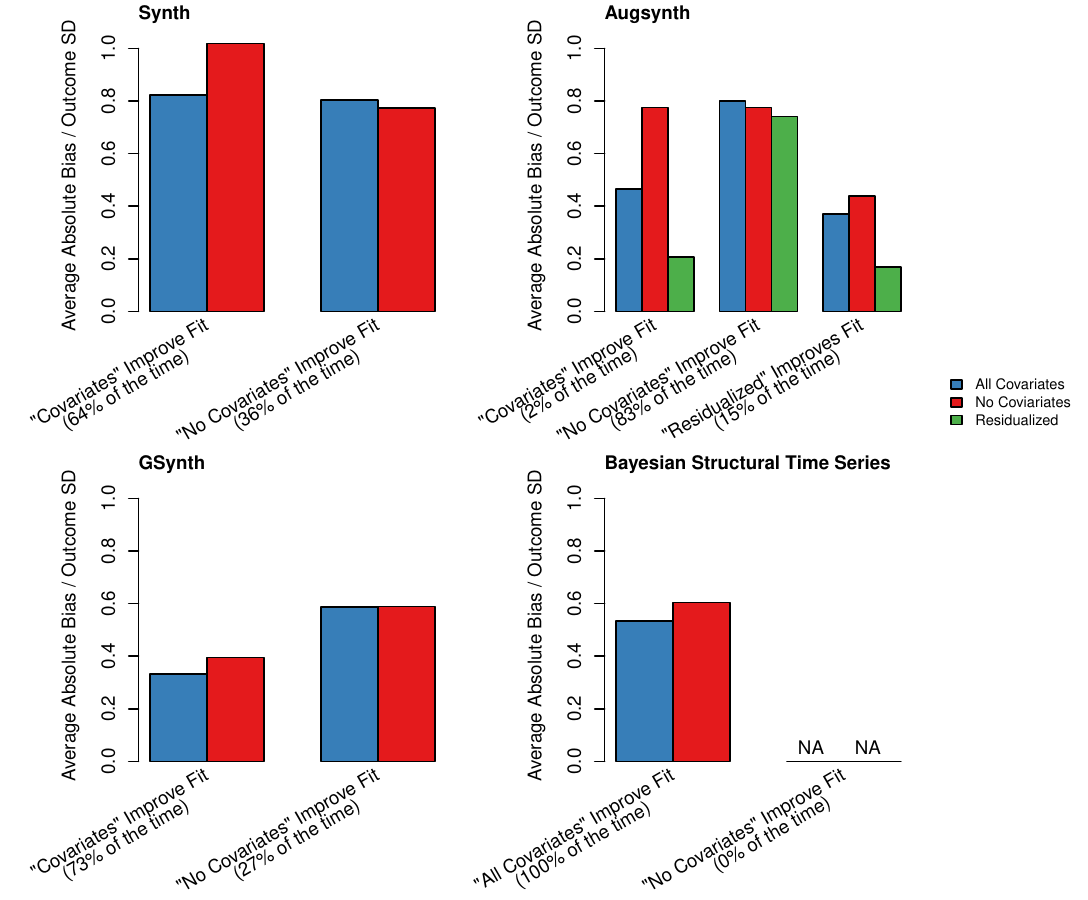}
\caption{\label{fig:fig4}The Effect of Covariates on Absolute Bias Conditional on Method Implementation that Produces the Lowest Imbalance.}
\end{figure}

Figure \ref{fig:fig4} presents these results for Synth, Augsynth, GSynth, and BSTS. In the Synth panel (top left), the first set of bars shows the average absolute bias when including all covariates improves pre-treatment outcome imbalance relative to excluding all covariates (all covariates improves imbalance ~64\% of the time). When including covariates improves pre-treatment outcome imbalance, including covariates also substantially reduces average absolute bias relative to excluding covariates. The second group of bars shows the results when excluding covariates improves pre-treatment outcome imbalance (36\% of the time). Here, as expected, excluding covariates also reduces average absolute bias relative to including covariates --- but only slightly. 

In the Augsynth panel, we see that excluding covariates most often results in lower pre-treatment outcome imbalance, but even when this is the case, residualization produced the lowest average absolute bias. In the cases where either `residualized' or `all covariate' implementations produces the lowest pre-treatment outcome imbalance, the `residualized' implementation still produces the lowest average absolute bias, followed by `all covariates' and finally `no covariates' produces the worst average absolute bias. For GSynth, including all covariates most often produces lower pre-treatment imbalance, and when it does, including all covariates also produces lower average absolute bias. When excluding covariates produces lower pre-treatment imbalance, the `all covariates' and `no covariates' implementations produce nearly identical average absolute bias. Finally, for BSTS, including all covariates always produced lower pre-treatment imbalance relative to excluding covariates in our simulations. Including all covariates in BSTS also reduced average absolute bias relative to the implementation that excludes covariates. 

In sum, these figures suggest that implementations that produce the lowest pre-treatment outcome imbalance are not necessarily the implementations that produce the lowest bias. This suggests that an \emph{over-reliance on pre-treament outcome imbalance may lead to overfitting and worse estimates.}

\subsection{Misconception 3: Lower Pre-Treatment Outcome Imbalance Suggests Lower Absolute Bias}
\begin{figure}[h]
\centering
\includegraphics[width=\textwidth,height=\textheight,keepaspectratio]{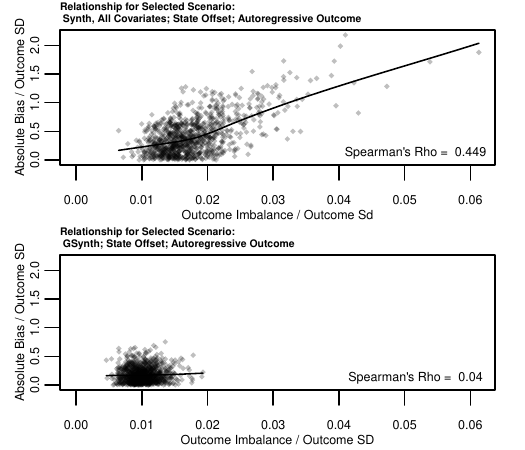}
\caption{\label{fig:fig5}Varying Relationship Between Pre-Treatment Outcome Imbalance and Absolute Bias Depending on Method and Scenario with Loess Line and Spearman's Rho.}
\end{figure}

We next turn to the relationship between pre-treatment outcome imbalance and absolute bias more generally. Building on the findings in Figure \ref{fig:fig4} that suggest that removing covariates to improve pre-treatment outcome imbalance is unlikely to reduce absolute bias, we generate scatterplots of pre-treatment outcome imbalance and absolute bias for each simulation scenario and method combination and calculate Spearman's rank correlation coefficient. Figure \ref{fig:fig5} provides two examples for the `state offset' simulation scenario with an autoregressive outcome model. The scatterplot for Synth with all covariates included looks as expected, with a relatively strong linear relationship between pre-treatment outcome imbalance and absolute bias (the resulting Spearman's Rho is 0.449). Interestingly, though, we do not find this relationship across all methods. Indeed, the scatterplot for GSynth is much less clear, with a resulting Spearman's Rho of 0.04.

\begin{figure}[h]
\centering
\includegraphics[width=\textwidth,height=\textheight,keepaspectratio]{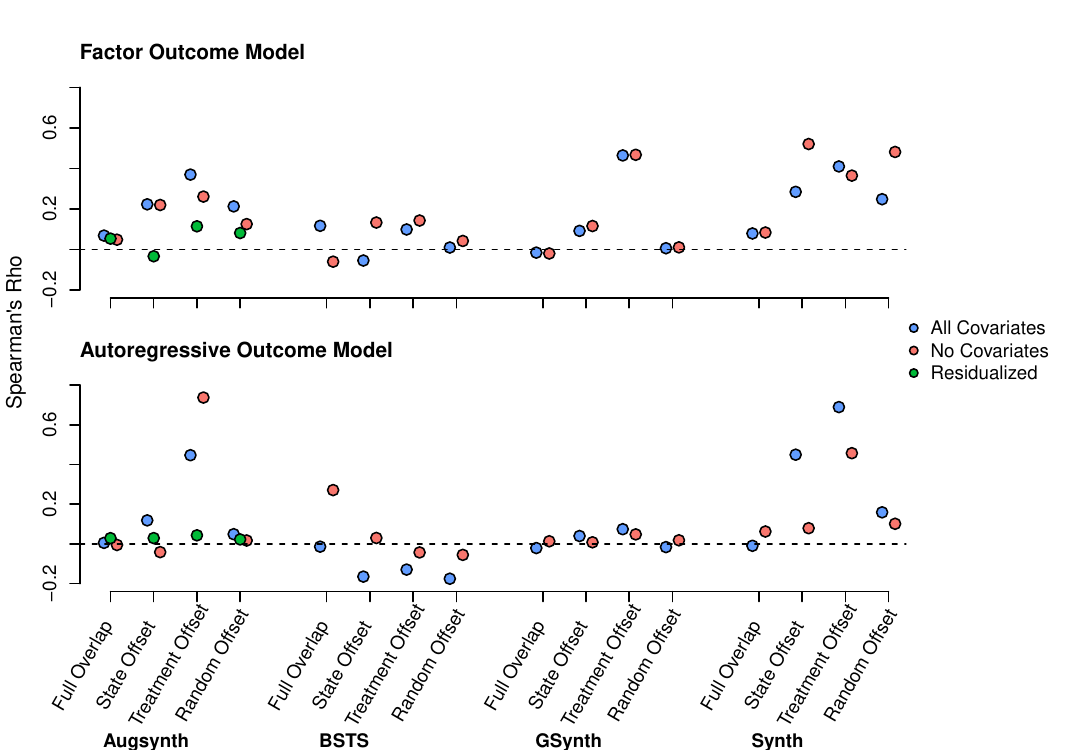}
\caption{\label{fig:fig6}Relationship (Spearman’s Rho) Between Pre-treatment Outcome Imbalance and Absolute Bias by Simulation Scenario, Method, Outcome Data Generating Model, and Inclusion of Covariates. Horizontal jitter added to help distinguish between points representing methods with similar performance.}
\end{figure}

Figure \ref{fig:fig6} displays Spearman's Rho for all method and simulation scenario combinations to explore this relationship holistically across our simulations. Surprisingly, we find that Spearman's Rho is often close to zero or even negative (especially for BSTS implementations). Notably, these relationships are not consistent across outcome data generating models. For example, GSynth methods with or without covariates in the `treatment offset' simulation scenario either has a reasonably strong relationship between pre-treatment outcome imbalance and absolute bias (factor outcome model) or almost no relationship (autoregressive outcome model). In this case, when the applied researcher calculates the pre-treatment outcome imbalance metric for their GSynth implementation, they won't have a way to ascertain whether or not that pre-treatment outcome imbalance is predictive of lower absolute bias or not.

\begin{figure}\textbf{}
\centering
\includegraphics[width=\textwidth,height=\textheight,keepaspectratio]{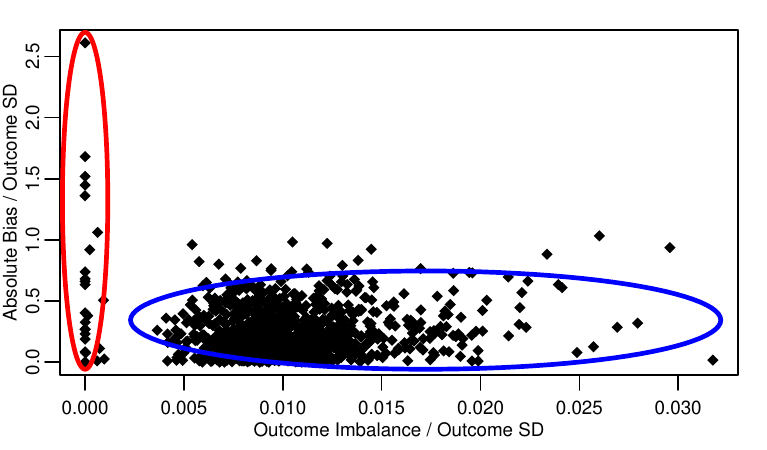}
\caption{\label{fig:fig7} Lower Outcome Imbalance Can Produce Higher Absolute Bias --- Augsynth with No Covariates, `Random Offset' Scenario with a Autoregressive Outcome Model.}
\end{figure}

Even for a single method within a single simulation scenario, the relationship between pre-treatment outcome imbalance and absolute bias can be ambiguous. Figure \ref{fig:fig7} shows the scatterplot of pre-treatment outcome imbalance and absolute bias for Augsynth without covariates in the `random offset' simulation scenario with an autoregressive outcome model. Here we see that for a majority of the 1,000 simulations there is a weak positive relationship between outcome imbalance and absolute bias (blue oval). There is an outlying cloud of points, however, with nearly perfect pre-treatment fit but with higher than average absolute bias (red oval). Applied researchers should thus be cautious of methods that produce nearly perfect pre-treatment fit --- they may in fact be too good to be true.

Finally, we consider the possibility that an applied researcher tries each method on a given data set, calculates the pre-treatment outcome imbalance, and wants to find the method that will produce the lowest absolute bias. To evaluate this situation, we rank each of our nine methods (Synth with and without covariates, GSynth with and without covariates, BSTS with and without covariates, Augsynth with covariates, without covariates, and residualized) by their pre-treatment outcome imbalance and absolute bias in each of our simulations for all of our simulation scenarios. 

\begin{figure}\textbf{}
\centering
\includegraphics[width=\textwidth,height=\textheight,keepaspectratio]{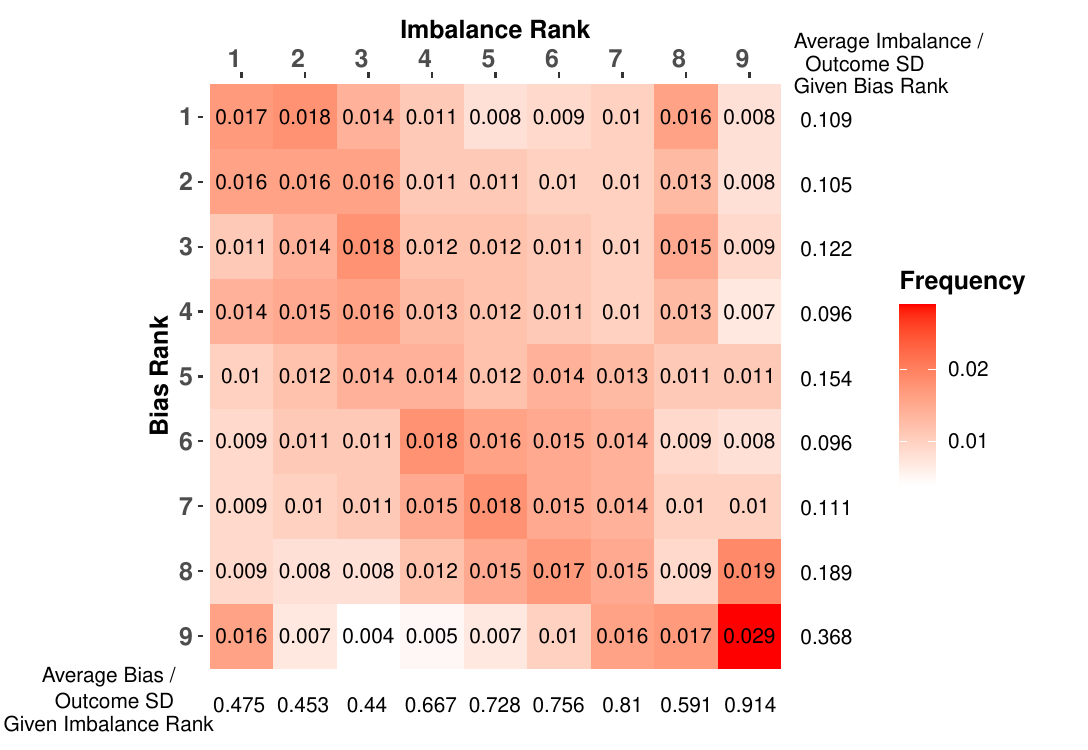}
\caption{\label{fig:fig8}Relationship Between Rank of Pre-Treatment Outcome Imbalance and Rank of Absolute Bias Across Methods.}
\end{figure}

We plot the proportions by imbalance and bias rank in Figure \ref{fig:fig8}. As indicated by the higher proportions surrounding the main diagonal, there is a general relationship between a method's pre-treatment outcome imbalance and that method's bias. There are some striking off-diagonal features, however, that suggest researchers should not simply select whichever method produces the lowest pre-treatment outcome imbalance. 

While the most common outcome when selecting the method that generates the lowest pre-treatment imbalance is that the method also produces the lowest absolute bias, the third most common outcome is that the method produces the worst absolute bias. Selecting the second-best method for pre-treatment imbalance finds the lowest absolute bias more often than the best pre-treatment imbalance method, and does not result in the highest absolute bias nearly as often. 

The summary statistics provided in the margins suggest that variance in average pre-treatment imbalance by rank in absolute bias is quite low --- methods in the top 7 out of 9 in terms of average absolute bias have very similar pre-treatment imbalance statistics (within 0.06 of a standard deviation of the outcome). Similarly, the top three methods in terms of imbalance have nearly identical average absolute bias (within 0.04 of an outcome standard deviation of each other). \emph{While lower pre-treatment outcome imbalance is generally suggestive of lower absolute bias, the relationship is weak and contains troubling outliers.}

\section{Discussion and Recommendations}

The accepted truths that dominate the application of synthetic control are actually myths. Throughout our simulation results, we consistently find evidence that these misconceptions are not consistent with empirical evidence. In this section we summarize our findings and share several recommendations for practice that are better supported than the misconceptions described above.

It is worth emphasizing that we do not expect the original authors of the SC technique to be surprised by these findings. Indeed many of the issues we describe here are echoes of their own recommendations. On the role of covariate weights, for instance, \citet[496]{abadie_synthetic_2010} notes, ``Although our inferential procedures are valid for any choice of $V$, the choice of $V$ inﬂuences the mean square error of the estimator.'' And in a paper detailing their own applied recommendations, \citet[400]{abadie_using_2021} warns against potential overfitting noting that ``In contrast, a small number of pre-intervention periods combined with enough variation in the unobserved transitory shocks may result in a close match for pretreatment outcomes even if the synthetic control does not closely match the values of $\mu_1$. This is a form of over-fitting and a potential source of bias.'' While our recommendations may not be new, we believe it is important to pair these statements with applied examples to contextualize how severe these issues can be. We hope our examples of overfitting in applied settings, for example, proves useful for those who might over-index on strong pre-treatment fit for short durations.

\subsection{Do not use regression weights}

\citet[pp. 496]{abadie_synthetic_2010} note that ``although [their] inferential procedures are valid for any choice of \textbf{V}, the choice of \textbf{V} influences the mean square error of the estimator." Our simulation results suggest that practitioners have paid too much attention to the first clause of this sentence and not enough attention to the second. While the choice of \textbf{V} weights may  be inconsequential asymptotically, in practical applications the choice appears to have clear and substantial effects on the performance of the estimator. Given the performance of each estimator, we see very little reason to rely on implementations of Synth with data driven \textbf{V} weights derived from regression. Researchers may want to consider implementations of Synth that rely on the nested optimization process \citep[though there may be other reasons to avoid the nested optimizer, e.g.,][]{kaul_synthetic_2015, malo_computing_2020}. Ideally, researchers would select Augsynth over either of these implementations, and in particular the residualized implementation of Augsynth.

\subsection{Consider whether you require localization; interactive fixed effects may suffice}

One of the selling points of SC is that the weighting component localizes comparisons to control units that are particularly similar to the treated unit \citep[e.g.,][]{arkhangelsky_synthetic_2021}. Our findings, however, suggest that in many cases the relatively simpler interactive fixed effects model implemented in GSynth may perform just as well (or better) than localized synthetic control approaches \citep[see also,][]{liu_practical_2022}. This also suggests that, while theoretically well motivated, the actual empirical benefits from synthetic control's localization may be rather limited. Some have argued that this localization may make SC a more credible causal estimate than comparable fixed effects models \citep[e.g.,][]{arkhangelsky_synthetic_2021}, but given the limited differences we see in performance across our simulations we are left skeptical of such claims. Additional research is needed to more fully explore the practical benefits of SC's localized weights in a broader array of applied settings in particular when there are subgroups of control units that vary in their similarity to the treated unit, or in situations where the true data generating processes are highly nonlinear.

\subsection{Use covariates when feasible}
As with Abadie et al.'s 2010 note about the \textbf{V} matrix, our results suggest that researchers have put too much emphasis on covariates being asymptotically irrelevant and have not focused enough on the significance of covariates in applied settings. We find that if covariates are relevant for the outcome of interest, it is often a good idea to include them in the synthetic control analysis --- even when excluding covariates improves pre-treatment outcome imbalance metrics. While there is considerable variation in the importance of covariates across simulations --- when excluding covariates reduces bias the gains appear to be quite moderate, but when including covariates reduces bias the gains appear more substantial. Given that researchers are unlikely to know the (unobserved) specifics of their data in order to know which situation they're in, the benefits of including covariates appear to outweigh the costs on average. An alternative is to include outcomes from \emph{all} pre-treatment time periods \citep{ferman_cherry_2020} which makes including additional covariates intractable \citep{kaul_synthetic_2015}. Ideally, researchers would consistently report the robustness of their findings across each of these specifications.

\subsection{Only omit reference categories if the choice is truly arbitrary}

If you include compositional variables among your covariates, be aware that standard methods are sensitive to the choice of which reference category you omit. The simplest solution to this problem is to instead use methods where the choice of reference category is truly arbitrary --- namely GSynth or residualized Augsynth. If neither of these methods are suitable, we recommend not omitting reference categories at all (e.g., \citet{degli_esposti_effects_2023}). Synth that includes all categories of compositional variables cannot initialize the \textbf{V} matrix with regression weights, but our results suggest that the nested optimizer initialized with uniform weights in the \textbf{V} matrix still does a reasonably good job. Augsynth (without residualization) and BSTS implementations rely on regularization for variable selection, and can handle collinear predictors without omitting a reference category. At the very least, results from Synth or Augsynth (without residualization) that omit reference categories should be interpreted as one of many plausible estimates, and researchers should incorporate that additional uncertainty when interpreting these results. Another approach is to provide all possible estimates to the reader.

\subsection{Don't rely heavily on pre-treatment outcome imbalance}

Our simulation results find remarkably little evidence that pre-treatment outcome imbalance is a reliable predictor of bias. While there may be a relationship in some method --- scenario combinations, that relationship does not hold generally. Indeed, for some methods pre-treatment outcome imbalance appears to be entirely unrelated to the bias of the predictor, regardless of the simulation scenario. While a strong pre-treatment outcome imbalance may be a necessary condition for producing a plausible synthetic control estimate, it does not appear to be a sufficient condition and results should be interpreted on their theoretical merits in addition to their pre-treatment goodness of fit.

Beyond within-method comparisons, we would caution against applied researchers using pre-treatment outcome balance as a metric to compare performance across methods. This comparison presumes a stable relationship between outcome imbalance and model performance which does not appear to hold in practice. For example, within the `treatment offset' simulation scenario with an autoregressive outcome model, GSynth might have a lower pre-treatment outcome imbalance than Synth, but the outcome imbalance is only predictive of bias in the case of Synth. Such comparisons are of apples to oranges, and cannot be used to identify the method with the lowest bias.

Finally, we would caution researchers that pre-treatment outcome imbalance can be too good to be true. In some method/scenario combinations a particularly low pre-treatment outcome imbalance is related with worse, not better, average bias. This is likely the result of over-fitting to the pre-treatment outcome series without enough attention to how that fit will (or won't) generalize post treatment.

\subsection{Limitations and Future Research}

Our simulations must represent real world scenarios for them to provide practical guidance. While we have attempted to provide a range of plausible scenarios that are informed by real world associations, we can make no claim that these simulations are representative of any real-world data generating process. Further, our set of simulations is in no way exhaustive of all possible real-world scenarios. Future research should explore how SC methods perform for a broader range of data generating processes, especially ones that are less tailored to SC methods (e.g., including non-linearities or processes that evolve over time). Our findings also suggest that particular methods may be more suited to some data generating processes than others. If researchers are able to identify which setting (approximately) underlies their data, they may be able to leverage the variation we identify to inform model selection. To do so, however, researchers would need to develop appropriate diagnostics to describe their data. We hope that such development will be a fruitful area for future research.

Finally, we have focused on pre-treatment RMSPE or 'outcome imbalance' because it is the diagnostic we have seen most commonly used in the literature. That said, it is certainly not the only diagnostic possible. Indeed, \citet{abadie_using_2021} suggest at least two other diagnostic and robustness checks. The first is backdating treatment, where SC methods are fit using an arbitrary treatment date some time before the actual treatment occurs. If the synthetic time series is a good approximation of the counterfactual time series, then the synthetic trajectory between the arbitrary and actual treatment dates should be a very close fit to the true data. The second is a leave-one-out test to evaluate the sensitivity to a particular donor unit or covariate included in the model. Future research is needed to evaluate how well these approaches (or some set of new diagnostics) predict model performance in practice --- either alone or in conjunction with pre-treatment RMSPE. Such work would hopefully clarify when these diagnostics are `good enough' to ensure reasonable estimates or when SC methods should be abandoned in particular applications.

\section{Conclusions}

The rapid rise in popularity of Synthetic Control methods have inspired a proliferation of methods, techniques, and implementations that have outpaced the literature evaluating these approaches. This asymmetry has left researchers to try to glean bits of wisdom either from the original proofs or from other authors description of the method and best practice in more applied papers to guide their practical applications. Some of the implications from the original theorems, however, may be inapplicable in applied settings and may cause some researchers to make decisions that are detrimental to their analyses. Moreover, some of the descriptions of the method and practical guidance in subsequent papers by other authors is either incorrect or incomplete. Our analysis reveals three such misconceptions where guidelines from theory lead researchers to sub-optimal conclusions --- the misconception that Synthetic Control is relatively insensitive to choices made in implementation, the misconception that covariates are unnecessary, and the misconception that pre-treatment outcome imbalance is predictive of better model performance.

We fear that underlying each of these misconceptions is an exaggerated notion of Synthetic Control's robustness. From that perspective, it is not surprising that incorporating additional information by including covariates that are related to the outcome often improves model performance. The settings in which covariates are asymptotically superfluous as the number of pre-treatment time periods increase require strict assumptions about the data generating process. Thus it is dangerous to conflate that mathematical result with general advice that covariates are typically irrelevant in practice. Similarly, conflating how well the synthetic control fits the pre-treatment trajectory with an assessment of how well the synthetic control represents the counterfactual appears overly optimistic at best.

The debunking of these misconceptions, and the relative comparison of performance across methods generally, suggest a caution against magical thinking when it comes to Synthetic Control. Although the method's theoretical precepts are appealing, the method does not (and should not) provide a silver bullet for all time-series causal questions. Indeed, we find that the much simpler interactive two-way fixed effects model performs just as well or better than synthetic control --- at least for these simulations. If this simpler model seems to less credibly identify causal effects, we would recommend that the more complicated synthetic control methods come under the same scrutiny. Though synthetic control methods can and do work well in some settings, we encourage researchers to approach such tools with the same optimistic trepidation they would bring to any of the more well trodden methodological techniques.   
\clearpage

\bibliography{SyntheticControl.bib}    
\bibliographystyle{apalike}

\clearpage
\begin{appendix}

\section{Simulation Specifics}\label{appendix:sim}

We first model each of the compositional covariates with Dirichlet regressions, which take the general form

\begin{equation*}
f(\boldsymbol{y}|\boldsymbol{\alpha}) = \dfrac{\Gamma (\sum_{i=1}^{k}\alpha_i)}{\prod_{i=1}^{k}\Gamma(\alpha_i)} \prod_{i=1}^k y_i^{\alpha_i - 1} 
\end{equation*}

and

\begin{gather*}
    ln\begin{bmatrix} \alpha_1 \\ \alpha_2 \\ \alpha_3 \\ \vdots \\ \alpha_k \end{bmatrix} = \begin{bmatrix} \mathbf{X_{s,t}}\boldsymbol{\beta}^1 + \lambda_s \\ \mathbf{X_{s,t}}\boldsymbol{\beta}^2 + \lambda_s \\ \mathbf{X_{s,t}}\boldsymbol{\beta}^3 + \lambda_s \\ \vdots \\ \mathbf{X_{s,t}}\boldsymbol{\beta}^n + \lambda_s \end{bmatrix}
\end{gather*}

where $k$ represents the number of mutually exclusive categories for the compositional covariates and the $\alpha$ parameters control the relative likelihood of each. These parameters are themselves a function of year and possibly the other covariates (with the exception of the first covariate which is modeled marginally).

The $\lambda_s$ parameter varies depending on the overlap scenario. The \emph{full overlap} scenario is the simplest scenario where we set $\lambda_s$ to be zero; that is, there are no fixed effects to create systematic differences between states. The \emph{treatment offset} scenario sets $\lambda_s$ to be a vector with an Alaska-specific fixed effect and zero for all other states. The \emph{state offset} scenario sets $\lambda_s$ equal to a state-specific fixed effect. In the \emph{random offset} scenario we model the empirical data just as in the fixed effects scenario, but randomly assign state intercepts for generating data - a process described in more detail below.

To calibrate our simulations to a realistic situation, we base many of the simulation components on estimates from fitted versions of the simulation models to the CPS-ASEC data. For the compositional variables we fit the following models to the CPS-ASEC data:

\begin{gather*}
    ln\begin{bmatrix} \alpha_{ind1} \\ \alpha_{ind2} \\ \alpha_{ind3} \\ \alpha_{ind4} \\ \alpha_{ind0} \end{bmatrix} = \begin{bmatrix} \beta^{ind1}_0 + \beta^{ind1}_1 * year_{s,y}  + \lambda_s\\ \beta^{ind2}_0 + \beta^{ind2}_1 * year_{s,y}   + \lambda_s \\ \beta^{ind3}_0 + \beta^{ind3}_1 * year_{s,y}  + \lambda_s \\ \beta^{ind4}_0 + \beta^{ind4}_1 * year_{s,y}   + \lambda_s \\ \beta^{ind0}_0 + \beta^{ind0}_1 * year_{s,y}   + \lambda_s \end{bmatrix}
\end{gather*}

Next, we fit models for the educational compositional variables:

\begin{small}
\begin{gather*}
    ln\begin{bmatrix} \alpha_{lths} \\ \alpha_{hs} \\ \alpha_{sc} \\ \alpha_{mtc} \end{bmatrix} = \begin{bmatrix} \beta^{lths}_0 + \beta^{lths}_1 * year_{s,y} +  \boldsymbol{\beta}^{lths}_{ind} * \boldsymbol{ind}_{s,y} + \lambda_s \\ \beta^{hs}_0 + \beta^{hs}_1 * year_{s,y} + \boldsymbol{\beta}^{hs}_{ind} * \boldsymbol{ind}_{s,y} + \lambda_s \\ \beta^{sc}_0 + \beta^{sc}_1 * year_{s,y} + \boldsymbol{\beta}^{sc}_{ind} * \boldsymbol{ind}_{s,y} + \lambda_s \\ \beta^{mtc}_0 + \beta^{mtc}_1 * year_{s,y} + \boldsymbol{\beta}^{mtc}_{ind} * \boldsymbol{ind}_{s,y} + \lambda_s \end{bmatrix}
\end{gather*}
\end{small}

Next, we fit models for the race compositional variables:
\small{
\begin{gather*}
    ln\begin{bmatrix} \alpha_{white} \\ \\ \alpha_{black} \\ \\ \alpha_{other} \end{bmatrix} = \begin{bmatrix} 
    \beta^{white}_0 + \beta^{white}_1 * year_{s,y} + \\
    \boldsymbol{\beta}^{white}_{ind} * \boldsymbol{ind}_{s,y}  + \boldsymbol{\beta}^{white}_{educ} * \boldsymbol{educ}_{s,y} + \lambda_s\\ \beta^{black}_0 + \beta^{black}_1 * year_{s,y} + \\ \boldsymbol{\beta}^{black}_{ind} * \boldsymbol{ind}_{s,y} + \boldsymbol{\beta}^{black}_{educ} * \boldsymbol{educ}_{s,y} + \lambda_s \\ \beta^{other}_0 + \beta^{other}_1 * year_{s,y} + \\ \boldsymbol{\beta}^{other}_{ind} * \boldsymbol{ind}_{s,y} + \boldsymbol{\beta}^{other}_{educ} * \boldsymbol{educ}_{s,y} + \lambda_s \end{bmatrix}
\end{gather*}}

Finally, we fit a linear regression for the wage variable:

\begin{eqnarray*}
    \begin{alignedat}{2}
wage_{s,y} & = \beta_0 + \beta_1 * year_{s,y} + \boldsymbol{\beta}_{ind} * \boldsymbol{ind}_{s,y} + \\
& \boldsymbol{\beta}_{educ} * \boldsymbol{educ}_{s,y} + \boldsymbol{\beta}_{race} * \boldsymbol{race}_{s,y}+ \lambda_s + \epsilon_{s,y} 
    \end{alignedat}
\end{eqnarray*}

We extract from these models estimates of the coefficients and variance matrices and insert them in the simulation models as shown below.

To generate simulated data, we start with an empty data set with each of the 50 states (plus D.C.) observed for years 1 through 100. We then draw coefficients for the industry model, letting $\hat{\boldsymbol{\eta}}$ be the vector of all estimated $\hat{\boldsymbol{\beta}}$ and $\hat{\lambda}_s$ parameters. We draw coefficients from $N(\boldsymbol{\hat{\eta}}, \boldsymbol{\hat{\Sigma}_\eta})$ where $\boldsymbol{\hat{\Sigma}_\eta}$ is the estimated parameter covariance matrix. We then plug our simulated year variable, $\Tilde{year}_{s,y}$ into the equation to generate predicted $\Tilde{\boldsymbol{\alpha}}_{ind}$ using our randomly drawn $\Tilde{\boldsymbol{\eta}}$ coefficients. Finally, we simulate the proportion in each of the five industry categories, $\Tilde{ind0}, \Tilde{ind1}, \ldots, \Tilde{ind4}$, by drawing from a dirichlet distribution using the generated $\Tilde{\boldsymbol{\alpha}}_{ind}$. We generate education compositional variables, $\Tilde{lths}, \Tilde{hs}, \Tilde{sc}, \Tilde{mtc}$, for each state and year in a similar manner, drawing coefficients from the education Dirichlet regression models and plugging in our simulated year and industry variables, then drawing simulated education compositional variables from a dirichlet distribution using the generated $\Tilde{\boldsymbol{\alpha_{educ}}}$. The same is true for race variables $\Tilde{white}, \Tilde{black}, \Tilde{other}$, where simulated year, industry, and educational variables are used as predictors. Finally, we generate the wage variable, $\Tilde{wage}$ by multiplying our simulated year, industry, education, and race covariates by coefficients drawn from $N(\boldsymbol{\hat{\eta}}, \boldsymbol{\hat{\Sigma}_\eta})$ where $\boldsymbol{\hat{\Sigma}_\eta}$ is the estimated parameter covariance matrix, and then adding noise from $N(0, \sigma^2_{\epsilon})$. We repeated this process of sampling new parameters and generating new covariates 1,000 times.

We generate four sets of 1,000 simulation datasets -- one for each overlap scenario -- by varying the nature of $\lambda_s$ in each simulation. The first three sets are straightforward manipulations --- 1) the \emph{full overlap} scenario sets $\lambda_s$ to zero; 2) the \emph{treatment offset} scenario sets $\lambda_s$ to be an Alaska-specific fixed effect and zero for all other states; and 3) the \emph{state offset} scenario sets $\lambda_s$ to be a state specific fixed effect. 

The \emph{random offset} scenario is somewhat more complicated. We start with the regression models as described for the \emph{state offset} scenario. To generate the data, however, we replace state fixed effects with randomly drawn state intercepts. Specifically, we draw intercepts from the distribution $N(0, (\dfrac{\sigma_\lambda}{3})^2)$ where $\sigma_\lambda$ is the standard deviation of the estimated state fixed effects from the empirical data. By randomly assigning state intercepts we allow for idiosyncratic differences between states within any given simulation dataset, but set the expectation of those differences to be zero across the full set of simulations. We divide $\sigma_\lambda$ by a factor of three in order to ensure that the resulting variability in simulated covariates is comparable to the variability of the empirical CPS-ASEC data.\footnote{The necessity for this adjustment likely stems from the random assignment of state intercepts breaking the covariance between estimated state fixed effects and the other estimated regression parameters.}

\subsection{Simulating Outcomes}

For each of the four overlap scenarios, we generate an outcome variable, conditional on the covariates, in one of two ways - with a linear model or with a factor model. As with the covariates, in order to calibrate our simulations to a real-life setting, we start by fitting models to CPS-ASEC data, regressing the proportion working part-time (our hypothetical outcome variable) on our compositional industry, education, and race variables as well as our continuous average wage variable.\footnote{Though the proportion working part-time is constrained to be within the range [0,1], we fit linear regressions for simplicity.} To avoid incorporating effects of the true PFD policy on the outcome, we only use CPS-ASEC data from years 1977 through 1981 to fit these outcome models. 

For the linear model, we fit:
\begin{equation*}
    \begin{alignedat}{2}
    pa&rt-time_{s,y} = \beta_0 + \beta_1 * year_{s,y} +\beta_2 * wage_{s,y} + \\
    &\boldsymbol{\beta}_{ind} * \boldsymbol{ind}_{s,y} + \boldsymbol{\beta}_{educ} * \boldsymbol{educ}_{s,y} + \\
    &\boldsymbol{\beta}_{race} * \boldsymbol{race}_{s,y}  + \epsilon_{s,y} 
    \end{alignedat}
\end{equation*} 
for $y \leq 1981$, with $\epsilon_{s,y}$ assumed to be normally distributed with mean 0. We extract estimated regression coefficients, $\mathbf{\hat{\beta}}$, as well as the variance covariance matrix, $\mathbf{\hat{\Sigma}_\beta}$ from this model.

As with the covariates, we then generate simulated outcome variables from a linear regression model with this structure. First we draw coefficients from $N(\boldsymbol{\hat{\beta}}, \boldsymbol{\hat{\Sigma}_\beta})$. Then we draw outcomes conditional on those parameters and our simulated model inputs (year, wage, industry, education, and race) from above.

For the factor model, we first calculate state-specific pre-treatment means for each of our covariates. We then regress the proportion working part time on year, these state-specific pre-treatment means, and the interaction between year and these pre-treatment means. Specifically, we fit the following model:

\begin{equation*}
    \begin{alignedat}{2}
pa&rt-time_{s,y} = \beta_1 * year_{y} + \beta_2 * \bar{wage}_{s} + \boldsymbol{\beta}_{ind} * \boldsymbol{\bar{ind}}_{s} + \\
&\boldsymbol{\beta}_{educ} * \boldsymbol{\bar{educ}}_{s} + \boldsymbol{\beta}_{race} * \boldsymbol{\bar{race}}_{s}  +\beta_3 * year_{y} * \bar{wage}_{s} + \\
&  \boldsymbol{\beta}_{ind}^{int} * \boldsymbol{\bar{ind}}_{s} * year_{y} + \boldsymbol{\beta}_{educ}^{int} * \boldsymbol{\bar{educ}}_{s} * year_{y} +\\
& \boldsymbol{\beta}_{race}^{int} * \boldsymbol{\bar{race}}_{s} * year_{y} + \epsilon_{s,y} 
\end{alignedat}
\end{equation*} 
for $y \leq 1981$, with $\epsilon_{s,y}$ assumed to be normally distributed with mean 0.

To generate simulated outcomes, we draw coefficients from $N(\boldsymbol{\hat{\beta}}, \boldsymbol{\hat{\Sigma}_\beta})$ and multiply these coefficients by our simulated year variable, and the state-specific pre-treatment means of our simulated wage, industry, education and race variables.
Finally, for each of the outcome variables, we add a constant treatment effect of about 0.017
in the state of Alaska 
for years 1977 through 1986.\footnote{This constant treatment effect represents an effect size of about 2 standard deviations. 
Since Synthetic Control Methods are fit using exclusively pre-treatment data, the magnitude of the treatment effect has no effect on the $\boldsymbol{w^*}$ weights, and, in turn, the estimated counterfactual trajectory.} 

In all, the two outcome models are crossed with the four covariate scenarios produce eight distinct simulation scenarios. To make comparisons across scenarios as clear as possible, covariate values are identical for each pair of outcomes within the four covariate scenarios.

\section{Variation by pre-treatment duration}\label{appendix:pre}

This appendix presents figures analogous to the ones presented in the main text, but for shorter pre-treatment time-series durations --- specifically 20 and 5 years. In general, results are markedly similar across durations, with a few notable differences specific to particular figures.

\subsection{Average Root Mean Squared Error of the Estimated Causal Effect by Simulation Scenario, Method, and Outcome Data Generating Model.}

Across the three pre-treatment durations, Augsynth continues to generally outperform Synth methods with either nested or regression weights, as measured by root mean squared error of the estimated causal effect. As pre-treatment durations get shorter, however, Augsynth's improvement relative to the Synth methods narrows. This is likely due to standard Synth methods (at least as implemented by default as they are here) balancing based on pre-treatment averages rather than the full time series. Thus, Synth may not be as effective as Augsynth at leveraging longer pre-treatment durations for improved model fit. On the flip side of that trade off, Augsynth may be overfitting to pre-treatment data when pre-treatment durations are short.

\begin{figure}[h]
\includegraphics[width=\textwidth,height=\textheight,keepaspectratio]{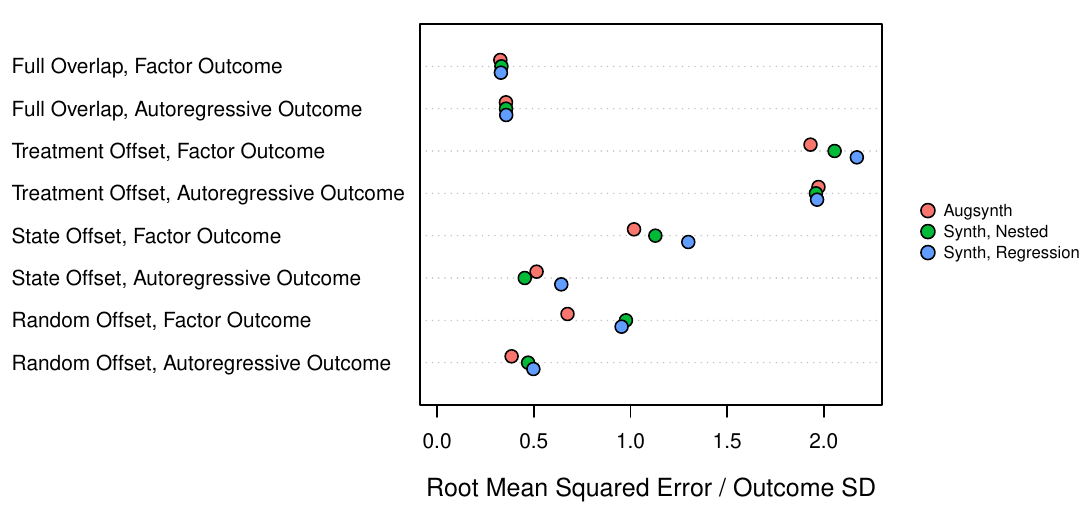}
\caption{\label{fig:figA20_1}Average Root Mean Squared Error of the Estimated Causal Effect by Simulation Scenario, Method, and Outcome Data Generating Model, 20 year pre-period.}
\end{figure}

\begin{figure}[h]
\includegraphics[width=\textwidth,height=\textheight,keepaspectratio]{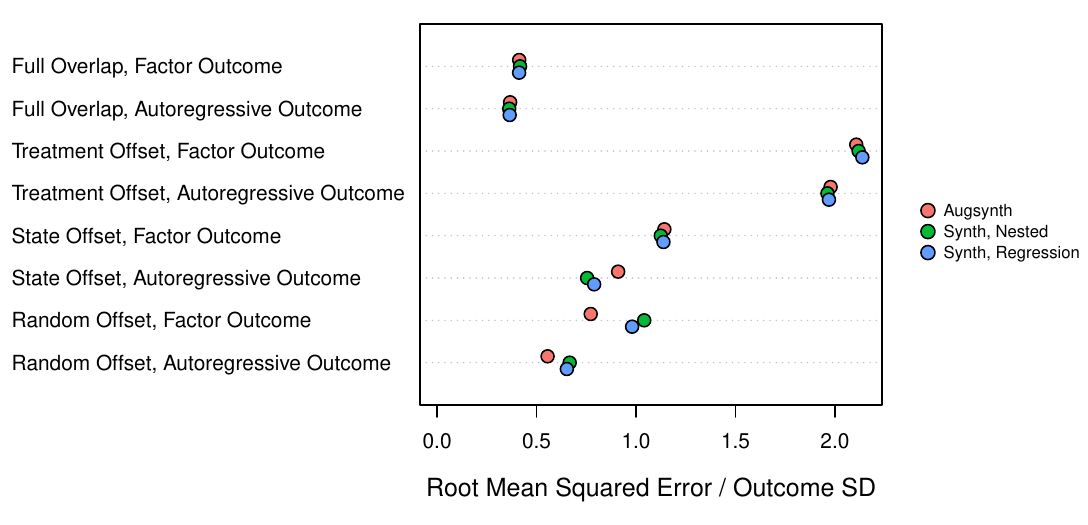}
\caption{\label{fig:figA5_1}Average Root Mean Squared Error of the Estimated Causal Effect by Simulation Scenario, Method, and Outcome Data Generating Model, 5 year pre-period.}
\end{figure}

\clearpage

\subsection{Variation in Synthetic Control Estimates Due To Reference Category Choice by Method and Simulation Scenario.}

Similarly to the finding for average root mean squared error of the estimated causal effect above, Augsynth tends to outperform Synth in uncertainty across reference categories when pre-treatment durations are long, but that performance decays as pre-treatment durations becomes short. In many scenarios, Augsynth underperforms relative to Synth in short pre-treatment durations in terms of reference category uncertainty --- perhaps reflecting Augsynth's potential for overfitting to short time-series.

\begin{figure}[h]
\includegraphics[width=\textwidth,height=\textheight,keepaspectratio]{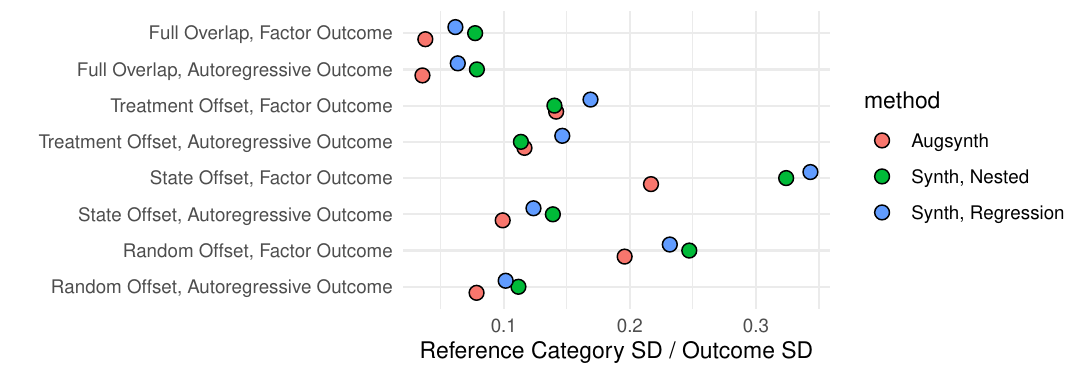}
\caption{\label{fig:figA20_2}Variation in Synthetic Control Estimates Due To Reference Category Choice by Method and Simulation Scenario, 20 year pre-period.}
\end{figure}

\begin{figure}[h]
\includegraphics[width=\textwidth,height=\textheight,keepaspectratio]{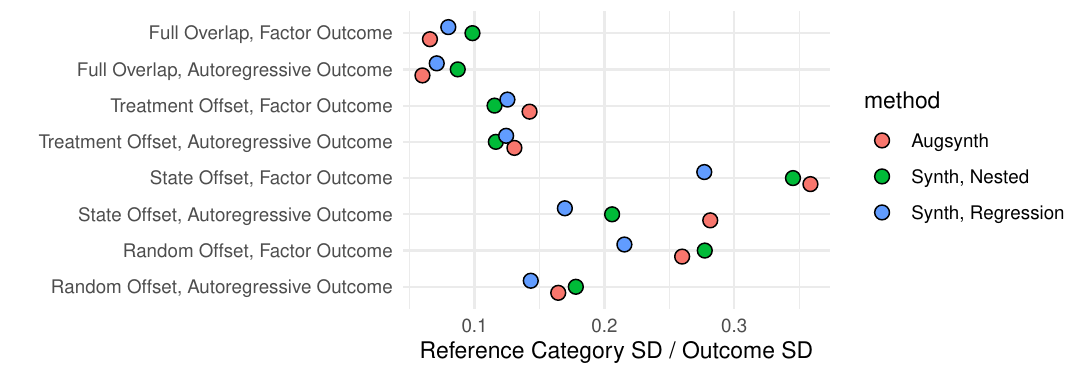}
\caption{\label{fig:figA5_2}Variation in Synthetic Control Estimates Due To Reference Category Choice by Method and Simulation Scenario, 5 year pre-period.}
\end{figure}

\clearpage

\subsection{Root Mean Squared Error of the Estimated Causal Effect Using All or No Covariates by Simulation Scenario, Method, and Outcome Data Generating Model.}

The relationship between the inclusion of covariates and RMSE of the estimated causal effect is ambiguous for long pre-treatment durations, and is increasingly ambiguous as pre-treatment durations get short. It appears that including covariates continues to result in slightly lower RMSE even in short pre-treatment durations, but those improvements are quite small.

\begin{figure}[h]
\includegraphics[width=\textwidth,height=\textheight,keepaspectratio]{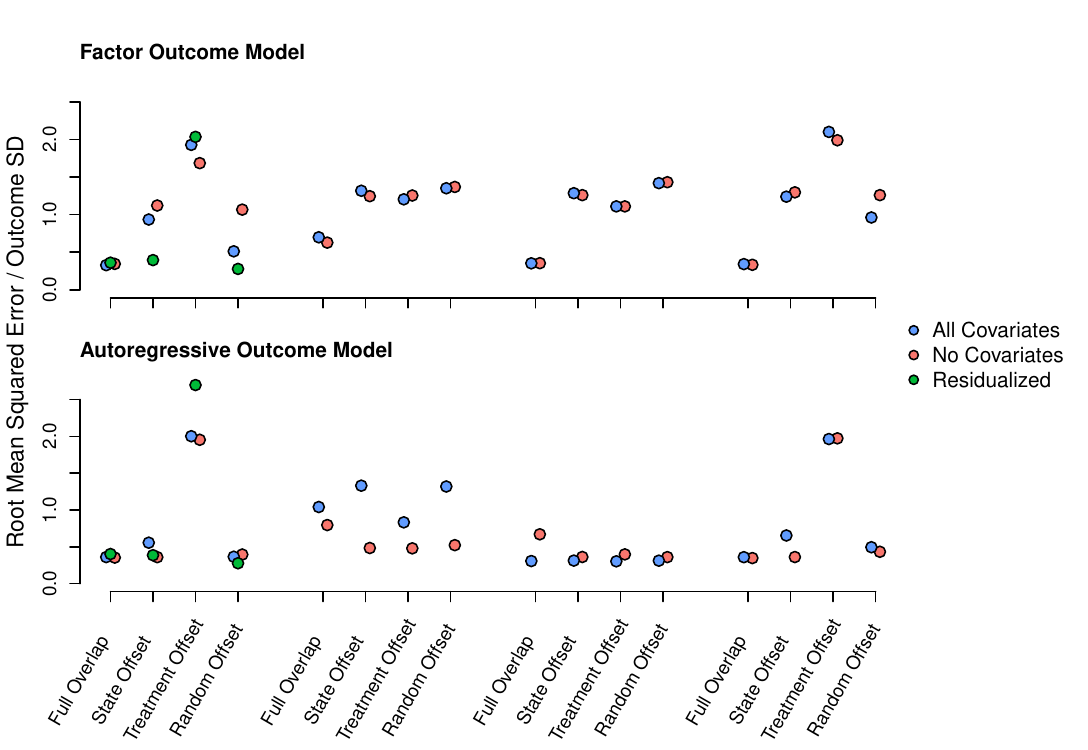}
\caption{\label{fig:figA20_3}Root Mean Squared Error of the Estimated Causal Effect Using All or No Covariates by Simulation Scenario, Method, and Outcome Data Generating Model, 20 year pre-period.}
\end{figure}

\begin{figure}[h]
\includegraphics[width=\textwidth,height=\textheight,keepaspectratio]{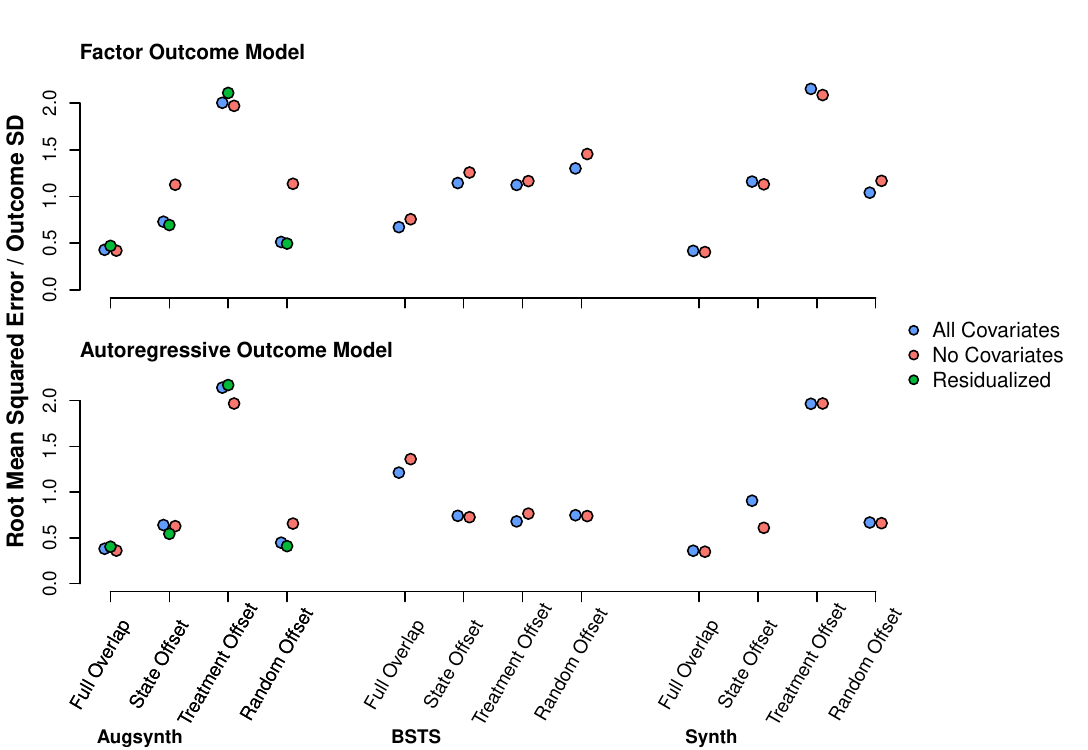}
\caption{\label{fig:figA5_3}Root Mean Squared Error of the Estimated Causal Effect Using All or No Covariates by Simulation Scenario, Method, and Outcome Data Generating Model, 5 year pre-period.}
\end{figure}

\clearpage

\subsection{The Effect of Covariates on Absolute Bias Conditional on Method Implementation that Produces the Lowest Imbalance.}

The relationship between the inclusion of covariates and average absolute bias conditional on whether including covariates improves pre-treatment outcome imbalance (RMSPE) appears relatively consistent across pre-treatment durations. It appears that including covariates tends to result in small improvements in average absolute bias across methods regardless of whether including covariates improves pre-treatment RMSPE.

\begin{figure}[h]
\includegraphics[width=\textwidth,height=\textheight,keepaspectratio]{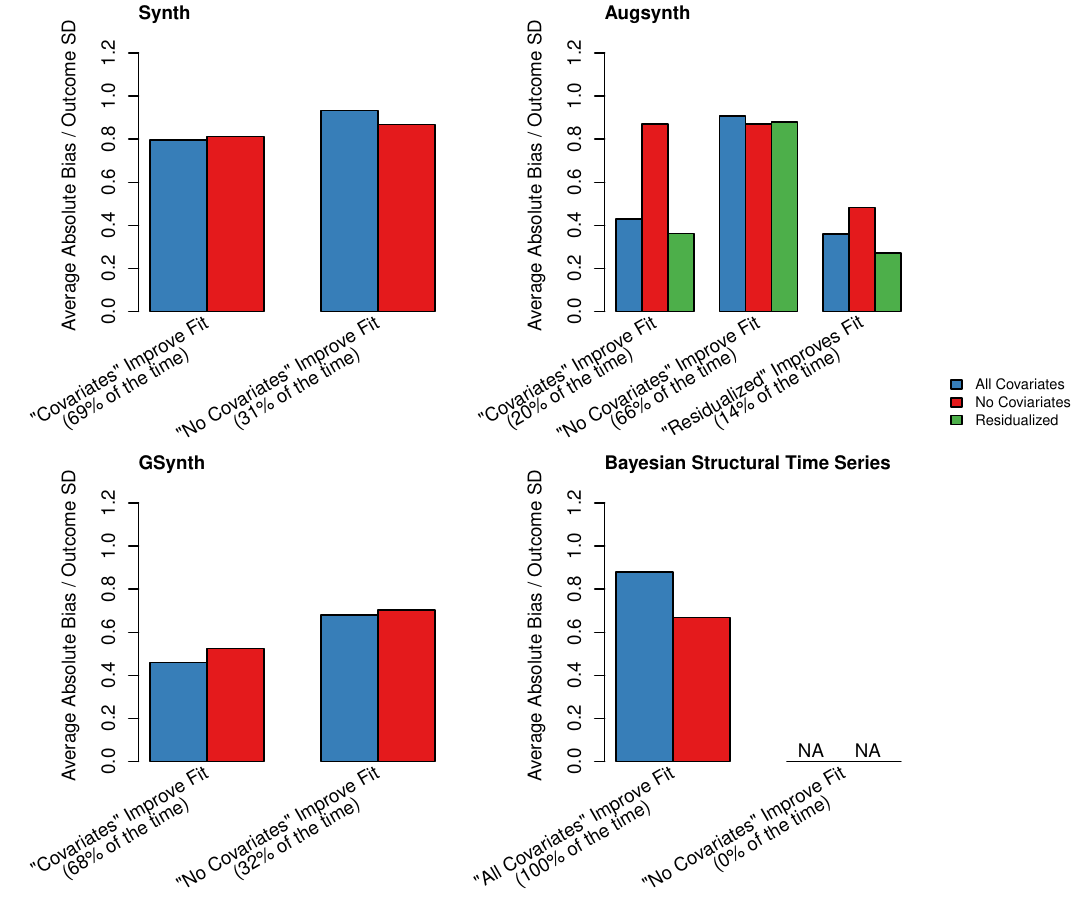}
\caption{\label{fig:figA20_4}The Effect of Covariates on Absolute Bias Conditional on Method Implementation that Produces the Lowest Imbalance, 20 year pre-period.}
\end{figure}

\begin{figure}[h]
\includegraphics[width=\textwidth,height=\textheight,keepaspectratio]{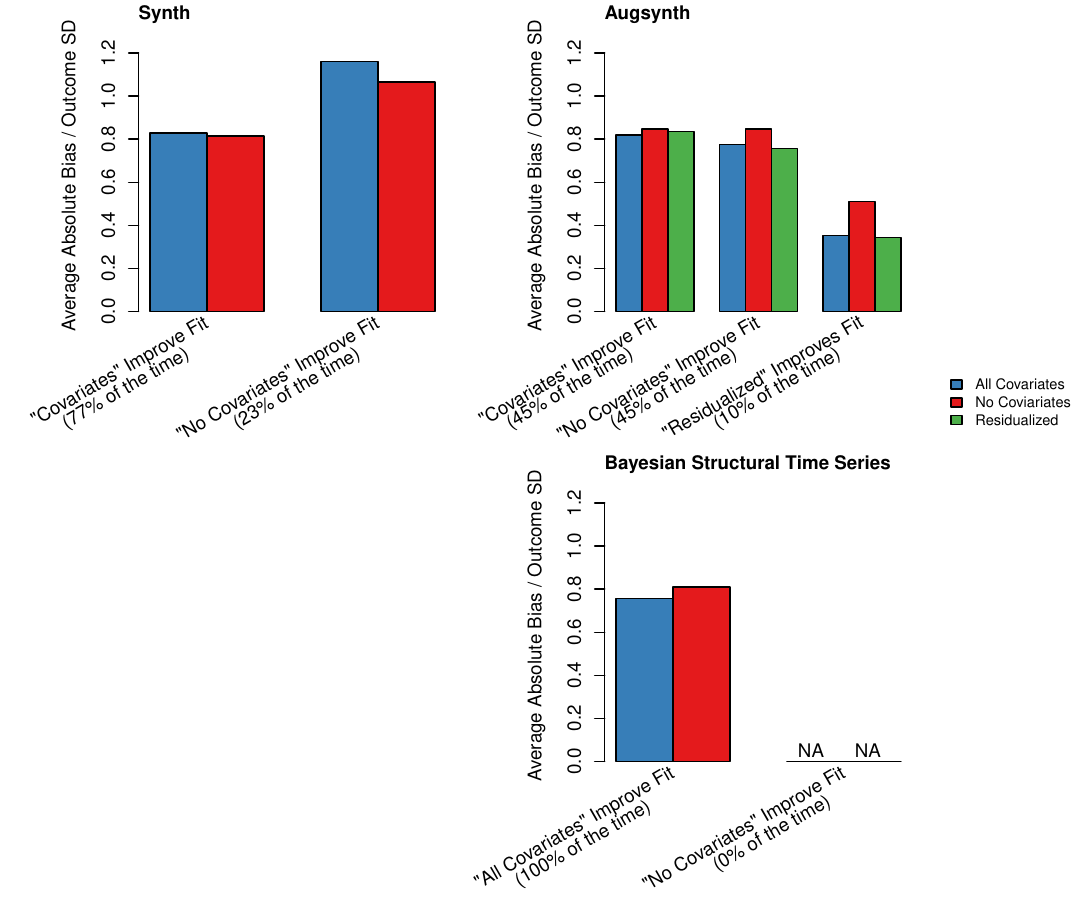}
\caption{\label{fig:figA5_4}The Effect of Covariates on Absolute Bias Conditional on Method Implementation that Produces the Lowest Imbalance, 5 year pre-period.}
\end{figure}

\clearpage

\subsection{Relationship Between Pre-treatment Outcome Imbalance and Absolute Bias by Simulation Scenario, Method, Outcome Data Generating Model, and Inclusion of Covariates.}

Across pre-treatment durations, the relationship between pre-treatment outcome imbalance (RMSPE) and average absolute bias is tenuous at best. The slight positive relationship observed for Synth methods in long pre-treatment duration declines as the pre-treatment duration declines.

\begin{figure}[h]
\includegraphics[width=\textwidth,height=\textheight,keepaspectratio]{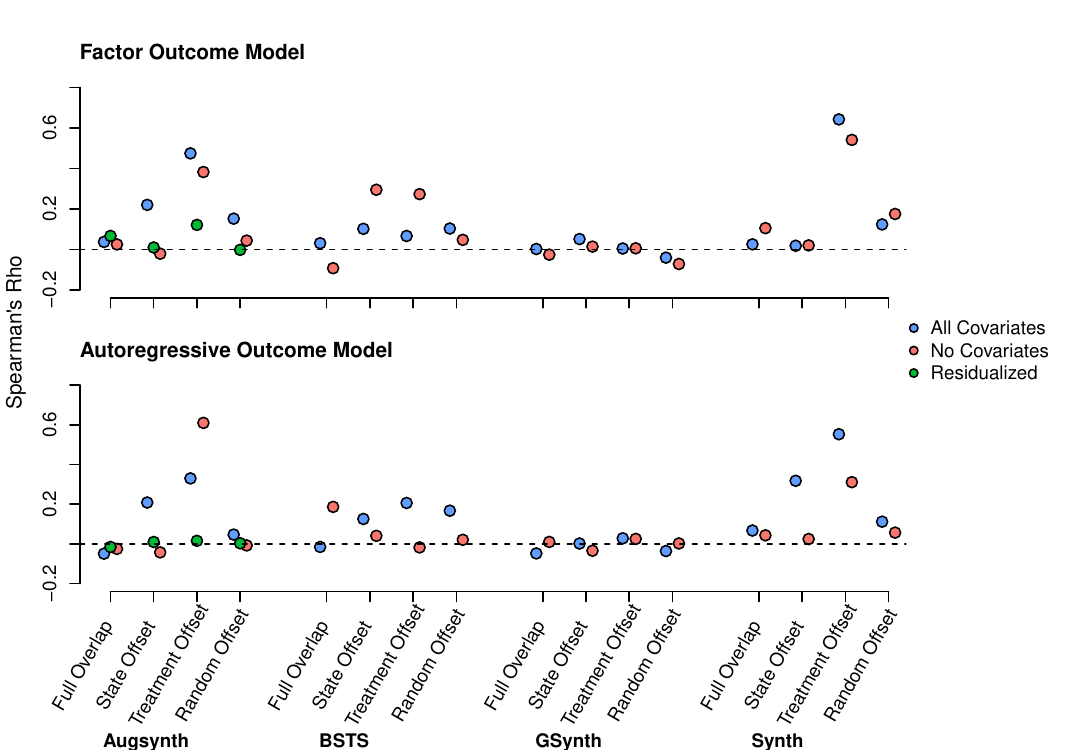}
\caption{\label{fig:figA20_6}Relationship (Spearman’s Rho) Between Pre-treatment Outcome Imbalance and Absolute Bias by Simulation Scenario, Method, Outcome Data Generating Model, and Inclusion of Covariates, 20 year pre-period.}
\end{figure}

\begin{figure}[h]
\includegraphics[width=\textwidth,height=\textheight,keepaspectratio]{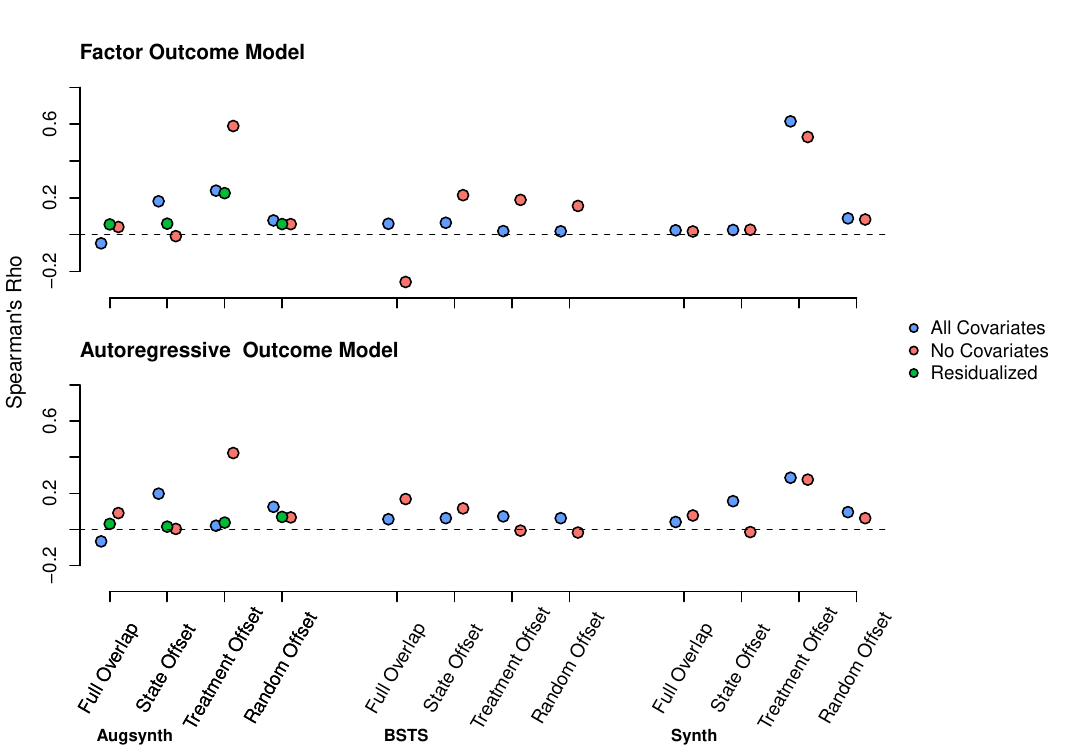}
\caption{\label{fig:figA5_6}Relationship (Spearman’s Rho) Between Pre-treatment Outcome Imbalance and Absolute Bias by Simulation Scenario, Method, Outcome Data Generating Model, and Inclusion of Covariates, 5 year pre-period.}
\end{figure}

\clearpage

\subsection{Relationship Between Rank of Pre-Treatment Outcome Imbalance and Rank of Absolute Bias Across Methods.}

The general relationship that methods which produce lower pre-treatment imbalance tend to produce lower average absolute bias appears to hold even as pre-treatment durations decline. Across all time periods, however, we see notable deviations from this general relationship and would caution researchers against choosing a method solely based on lower pre-treatment outcome imbalance (RMSPE).

\begin{figure}[h]
\includegraphics[width=\textwidth,height=\textheight,keepaspectratio]{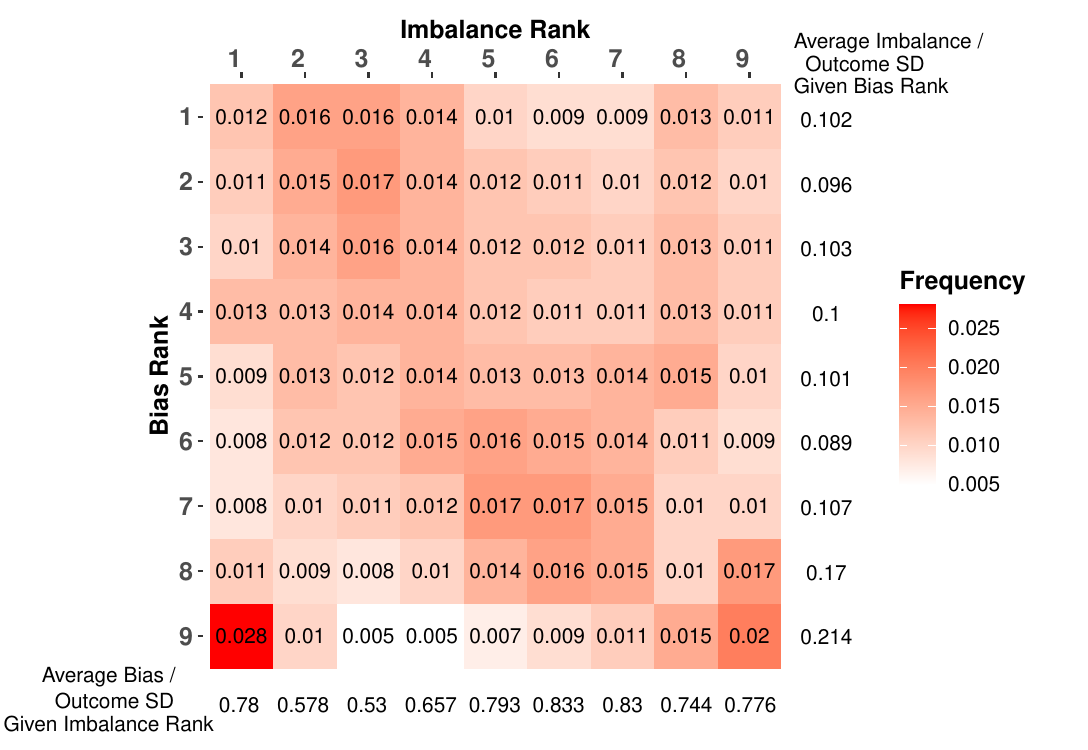}
\caption{\label{fig:figA20_8}Relationship Between Rank of Pre-Treatment Outcome Imbalance and Rank of Absolute Bias Across Methods, 20 year pre-period.}
\end{figure}

\begin{figure}[h]
\includegraphics[width=\textwidth,height=\textheight,keepaspectratio]{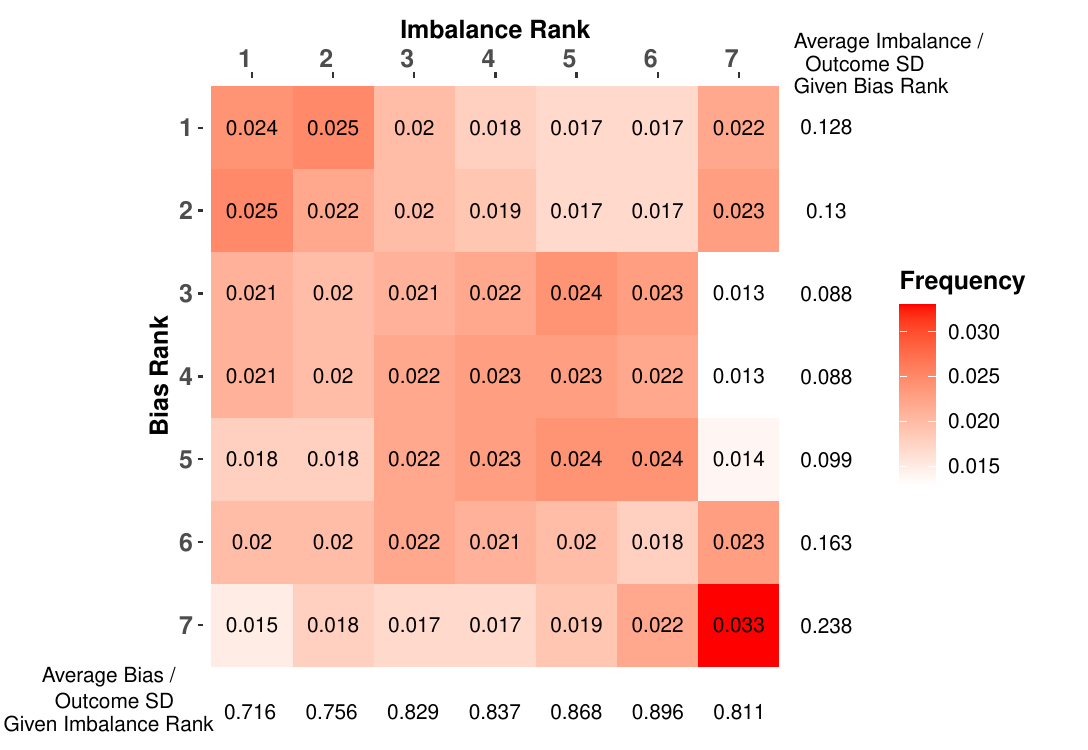}
\caption{\label{fig:figA5_8}Relationship Between Rank of Pre-Treatment Outcome Imbalance and Rank of Absolute Bias Across Methods, 5 year pre-period.}
\end{figure}
\end{appendix}

\end{document}